\documentclass{aa}
\usepackage{graphicx,fixltx2e}
\usepackage{amsmath,hyperref,color}

\usepackage{natbib} 
\bibpunct{(}{)}{;}{a}{}{,}
\begin{document}

 \title{Inferring telescope polarization properties through spectral lines without linear polarization}

   \author{A. Derks\inst{1,2} \and C. Beck\inst{2} \and V. Mart{\'i}nez Pillet\inst{2} }
        
   \titlerunning{Inferring telescope polarization through lines without linear polarization} 
  \authorrunning{A. Derks et al.}  
\offprints{A. Derks}

   \institute{University of Colorado, Boulder\\
     \and National Solar Observatory}
 \date{Received xxx; accepted xxx}
\abstract{Polarimetric observations taken with ground- or space-based
telescopes usually need to be corrected for changes of the polarization state
in the optical path.}{We present a technique to determine the polarization
properties of a telescope through observations of spectral lines that have no
or negligible intrinsic linear polarization signals. For such spectral lines,
any observed linear polarization must be induced by the telescope optics. We
apply the technique to observations taken with the Spectropolarimeter for
Infrared and Optical Regions (SPINOR) at the Dunn Solar Telescope (DST) and
demonstrate that we can retrieve the characteristic polarization properties of
the DST at three wavelengths of 459, 526, and 615\,nm.}{We determine the amount
of crosstalk between the intensity Stokes $I$ and the linear and circular
polarization states Stokes $Q$, $U,$ and $V$, and between Stokes $V$ and Stokes
$Q$ and $U$ in spectropolarimetric observations of active regions. We fit a set
of parameters that describe the polarization properties of the DST to the
observed crosstalk values. We compare our results to parameters that were
derived using a conventional telescope calibration unit (TCU).}{The values for
the ratio of reflectivities $X = r_s/r_p$ and the retardance $\tau$ of the DST
turret mirrors from the analysis of the crosstalk match those derived with the
TCU within the error bars. We find a negligible contribution of retardance from
the entrance and exit windows of the evacuated part of the DST. Residual crosstalk after applying a correction for the telescope polarization stays at a level of 3-10\,\% regardless of which parameter set is used, but with an rms fluctuation in the input data of already a few percent. The accuracy in the determination of the telescope properties is thus more limited by the quality of the input data than the method itself.}{It is possible to derive the parameters that describe
the polarization properties of a telescope from observations of spectral lines
without intrinsic linear polarization signal. Such spectral lines have a dense
coverage (about 50\,nm separation) in the visible part of the spectrum
(400--615\,nm), but none were found at longer wavelengths. Using spectral lines
without intrinsic linear polarization is a promising tool for the polarimetric
calibration of current or future solar telescopes such as the Daniel K.~Inouye
Solar Telescope (DKIST).} 
\maketitle
\section{Introduction \label{sec:introduction}}
Most existing solar telescopes were designed without clear requirements for
their polarimetric performance and calibration. The telescopes currently
available for high-spatial-resolution solar physics are calibrated using an analytical model of their time-varying Mueller matrices. The free
parameters in these models are fit to data obtained using calibration
polarizers at the entrance of the telescopes
\citep[][]{skumanich+etal1997,beck+etal2005a,selbing2010,socasnavarro+etal2011}
or used literature values for the polarization properties of the mirrors
\citep{collados1999,schlichenmaier+collados2002}.

The calibration strategy of the Advanced Stokes Polarimeter
\citep[ASP;][]{elmore1992}  at the Dunn Solar Telescope
\citep[DST][]{dunn1964,dunn+smartt1991} as described in
\citet[][]{skumanich+etal1997} paved the way for other telescopes to carry out
similar polarimetric characterizations. A key aspect of the ASP/DST calibration
was to separate the polarization effects of the telescope with
a time-dependent Mueller matrix ${\bf T}(t)$ from those of the
polarimeter with a static Mueller matrix ${\bf X}$ by inserting
polarization calibration optics between the two. This point of insertion
made the polarimeter matrix ${\bf X}$ constant in time but made the telescope Mueller
matrix ${\bf T}$ time-dependent to account for the varying telescope geometry.
The calibration optics provided the polarization properties of all of the
optical components downstream from the insertion point, i.e., the polarimeter
matrix. Upstream from the insertion point, the data obtained with the entrance
polarizer constrained the polarization properties described by the matrix
${\bf T}(t)$.  Polarization calibration optics usually have a small diameter and can be
manufactured to high accuracy compared to the large optics used for
entrance polarizers. Placing such high-quality calibration optics as high
up in the optical path as feasible is thus beneficial.

The French T\'elescope H\'eliographique pour l'Etude du Magn\'etisme et
des Instabilit\'es Solaires \citep[THEMIS;][]{lopezariste+etal2000} and the
proposed Large Earth-based Solar Telescope \citep[LEST;][]{engvold1991}
followed a different strategy for their polarization calibration. They minimized instrumental polarization effects by mounting
the polarization analyzer upstream of any oblique reflection. This design effectively transformed these telescopes into dedicated polarimeters. However, placing the analyzer so high up in the optical beam has several drawbacks. THEMIS uses a dual-beam scheme which implies the transfer of two linearly polarized beams through various optical elements
which leads to different optical aberrations in them.
LEST was designed to use only one of the linearly polarized beams with 
a loss of 50\% of the photons. Without a dual-beam scheme to
minimize the spurious seeing-induced polarization signals
\citep[][]{lites1987}, LEST was forced to rely on  high-speed detectors of
the Z{\"u}rich Imaging Polarimeter type \citep[ZIMPOL;][]{stenflo1985}.
We note also that the presence of entrance windows in both telescopes
prevented them from being completely free of instrumental polarization effects. 

In light of these circumstances, all modern solar telescopes adopted the
strategy of the ASP and incorporated calibration optics as high up in the
optical path as possible, thus calibrating the largest number of optical
components. One recent example is the German GREGOR telescope
\citep[][]{schmidt+etal2012} that uses a polarization calibration unit near the
secondary focus of the telescope, right behind the primary and secondary
mirrors. The calibration unit is located before breaking the revolution
symmetry around the optical axis. Thus, the expected polarization effects from
the mirrors upstream are small, originating only from the off-axis configuration outside the
center of the field of view and diffraction effects.
\citet{sanchezalmeida1992} demonstrated that these effects are only
important for absolute polarimetry at the $10^{-4}$ level. For GREGOR,
the matrix downstream of the calibration optics is geometry- and
time-dependent, whereas the matrix upstream is constant in time, and, in this
case, close to the identity matrix.

The Daniel K. Inouye Solar Telescope \citep[DKIST;][]{McMullin+etal2016}
under construction on the island of Maui will
follow a similar strategy as GREGOR. A polarization calibration unit is
located downstream of the secondary mirror. In the case of DKIST, the
first two mirrors M1 and M2 have an off-axis configuration that generates
a telescope matrix ${\bf T}$ that deviates significantly from the identity 
matrix, and is constant in time. The angles of incidence on these
two mirrors vary between 6 and 20 degrees. Accurate ray-tracing with a realistic modeling of the mirror coatings demonstrates that $U\leftrightarrow V$ crosstalk terms of the order of 8\% can be expected in ${\bf T}$
\citep[][]{harrington2016,harrington2017}. Interestingly, the same calculations show that the
diagonal elements are all close to unity to within less than a percent. 
Like for GREGOR, all temporal dependencies are transferred to the 
matrix downstream of the calibration optics. The polarization
properties of ${\bf T}$ are independent of the telescope pointing
and change only with the degradation of the coatings with time. 
Thus, the inference of this matrix is much simpler than in the case of the DST, the Swedish Solar Telescope (SST) 
or the German Vacuum Tower Telescope (VTT). 

The 4-m aperture of DKIST prevents usage of an entrance linear
polarizer. A number of strategies have been devised to calibrate the
instrumental polarization to the required level of accuracy of $5 \times 10^{-4}$
\citep[see][]{socasnavarro+etal2005d,socasnavarro2005e,socasnavarro2005f,elmore2014}.
One option for the calibration of the DKIST ${\bf T}$-matrix is
based on spectral lines that have no intrinsic linear polarization
through the Zeeman effect: for example, the \ion{Fe}{ii} line at 614.9\,nm
\citep[][]{lites1993}. A list of lines with this property, which
depends on the atomic transitions involved, was compiled by
\citet[][]{VillahozandAlmeida+1993} and \citet[][]{VillahozandAlmeida+1994}.
These authors proposed the use of those lines to measure the asymmetries of Stokes $V$ profiles as they are unaffected by instrumental polarization.

In this paper, we revisit their list of lines without linear
polarization to study their suitability for polarimetric calibration. We
compare the results obtained from observations of lines without linear
polarization taken in 2016 at the DST to those from the standard calibration done in 2010 using the
telescope entrance linear polarizers to understand the limitations of the
approach. The calibration of the ${\bf T}(t)$ matrix for the DST is more
complex than the calibration of ${\bf T}$ for DKIST, thus succeeding with
the DST demonstrates the potential interest of this method for DKIST. 

Section \ref{sec:synthesis} describes our selection process to identify
suited spectral lines. Section \ref{sec:application} gives the theoretical
background for the derivation of telescope properties from observations. The
observations used are described in Sect.~\ref{sec:observations}. Sections
\ref{sec:data analysis} and \ref{sec:results} provide the data analysis and
results, respectively. Section \ref{sec:discussion} contains the discussion,
while Sect.~\ref{sec:conclusions} gives our conclusions. 
The appendices provide the complete line list (Appendix \ref{appa}), a few more data examples at other wavelengths (Appendix \ref{app_transi}) and error estimates of crosstalk measurements (Appendix \ref{errestim}). 

\section{Synthesis of spectral lines and line selection \label{sec:synthesis}}
There exist special atomic transitions that produce spectral lines without any, or with negligible, intrinsic linear polarization (LP) induced by the Zeeman effect \citep{VillahozandAlmeida+1993,VillahozandAlmeida+1994}. These lines can be used to uncover polarization properties of a telescope or to test the quality of the polarimetric calibration because most, or all, linear polarization signal is spurious and indicates polarization crosstalk. We analyzed  the two tables of spectral lines without or with small LP provided by \citet[][their Table 1 and 2]{VillahozandAlmeida+1994} to find the lines that would be  best suited for telescope calibration purposes. We restricted the analysis to lines with rest wavelengths in the range 400--1100\,nm. We obtained all necessary information including excitation potential, log(gf), transition parameters, and the effective Land\'e coefficient (g\textsubscript{eff}) from the NIST Atomic Spectral Database\footnote{\url{http://physics.nist.gov/PhysRefData/ASD/lines_form.html}}. We disregarded any lines for which the value of log(gf) could not be found. This left 31 lines from Table 1 of \citet{VillahozandAlmeida+1994} (see Table \ref{table_wolp}) and 56 lines from their Table 2 (see Table \ref{table_wslp}).

After the suitable lines were chosen, a spectral synthesis was done using the Stokes Inversion based on Response functions code \citep[SIR;][]{ruizcobo+deltoroiniesta1992}. For this synthesis, we used a solar model atmosphere with a magnetic field inclination of $45\,^{\circ}$ and an azimuth of $22.5\,^{\circ}$, a macro- and micro-turbulent velocity of 1 km\,s$^{-1}$, and zero Doppler shift. The magnetic field inclination and azimuth were chosen in order to obtain polarization signal in all Stokes parameters $Q$, $U$, and $V$, whereas the macro- and micro-turbulence were selected so that realistic polarization amplitudes would be obtained. The synthesis was repeated with three different magnetic field strengths: 100\,G (quiet Sun), 1500\,G (plage), and 2500\,G (sunspot) to simulate regions of different magnetic activity on the solar surface. 
\begin{figure}
 \centerline{\resizebox{8.8cm}{!}{\includegraphics{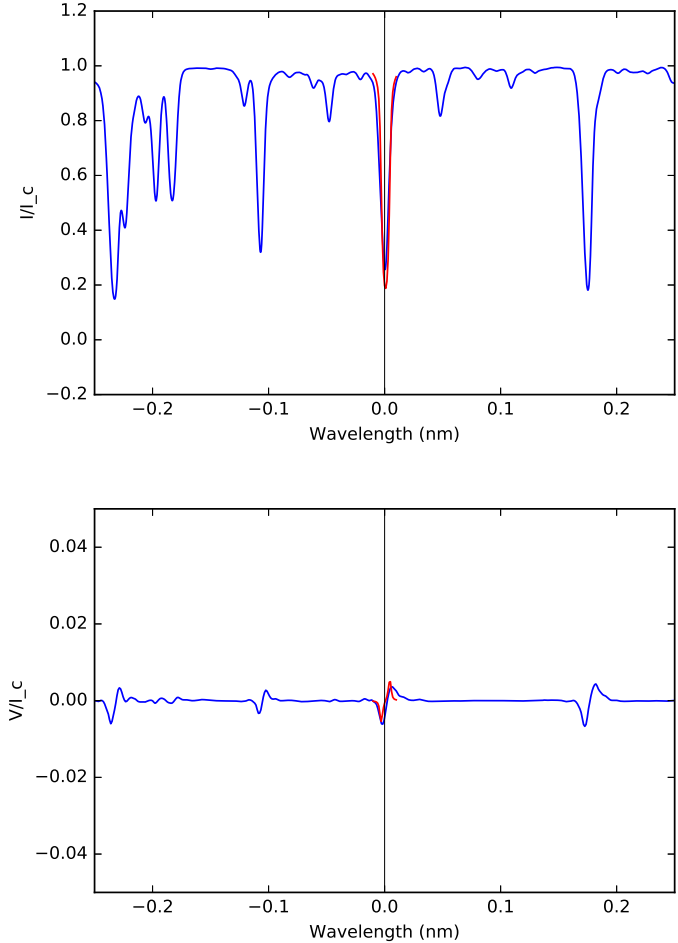}}}
\caption{Synthetic Stokes profiles of Stokes $I$ (top panel) and $V$ (bottom panel) for the \ion{Cr}{ii} line at 458.820\,nm in the synthesis with 1500\,G shown in red. The blue lines show the corresponding profiles from the FTS atlas. \label{fig1}}
\end{figure}
\begin{figure}
   \centerline{\resizebox{8.8cm}{!}{\includegraphics{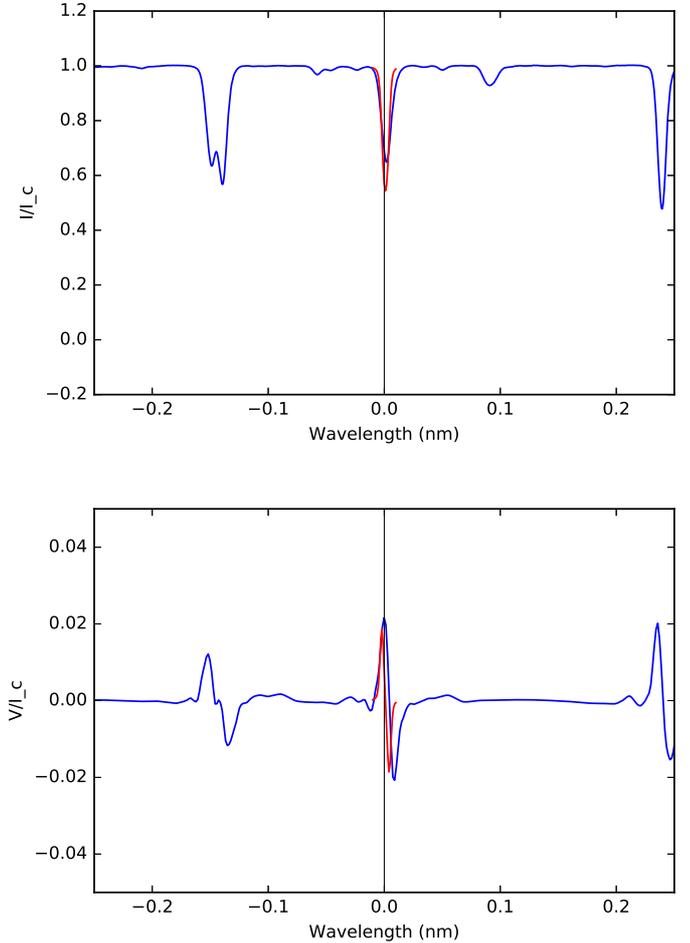}}}
  \caption{As in Fig.~\ref{fig1} but for \ion{Fe}{ii} at 614.923\,nm.\label{fig2}}
\end{figure}
\begin{figure*}
  \centerline{\resizebox{17.6cm}{!}{\includegraphics{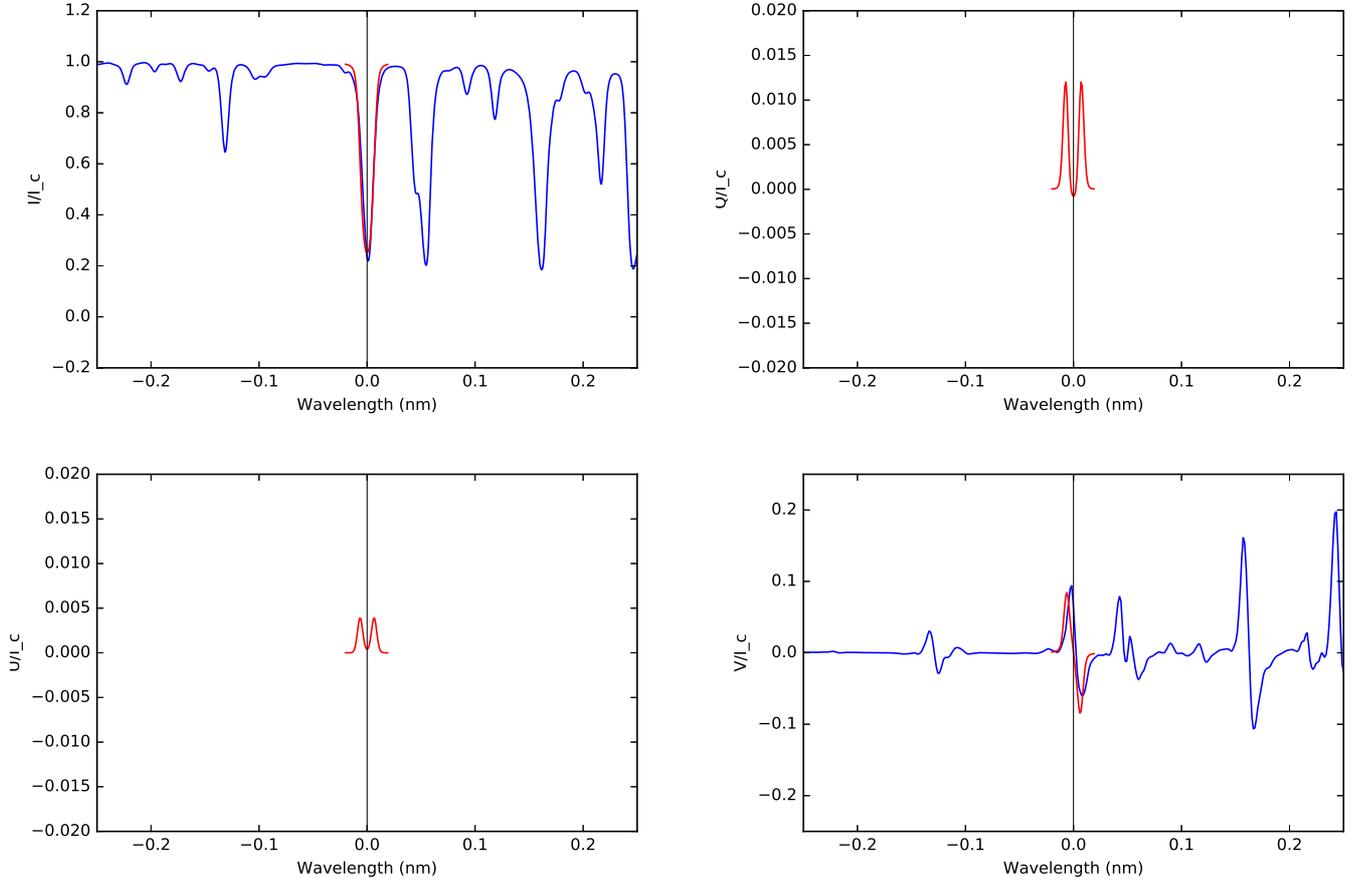}}}
  \caption{Synthetic Stokes profiles of Stokes $I$ (top left), $Q$ (top right), $U$ (bottom left) and $V$ (bottom right) for the \ion{Ca}{i} line at 526.171\,nm in the synthesis with 1500\,G. There is no corresponding FTS atlas data with Stokes Q and U measurements.\label{fig3}}
  \end{figure*}

Using the synthetic profiles, we made a final selection of lines suitable for calibration purposes based on line depth, polarization amplitude, and absence of line blends. Line blends were identified by a comparison to spectra from the Fourier Transform Spectrometer \citep[FTS;][]{kurucz+etal1984} atlas. 
We extracted the line depth from the 100\,G synthesis, and the maximal Stokes $Q$, $U$, and $V$ amplitudes for all lines from the 1500\,G synthesis (see Tables \ref{table_wolp} and \ref{table_wslp}).  
We identified a subset of six lines without or with negligible linear polarization (see Table \ref{tab1}) for testing the approach of inferring polarimetric telescope properties. 
These lines were selected because of their large line depths and polarization amplitudes, the absence of line blends (cf.~Figs.~\ref{fig1}--\ref{fig3}, Figs.~\ref{fig_synth5}--\ref{fig_synth6}), and a continuous wavelength coverage from 400 to 615\,nm. Figures \ref{fig1} to \ref{fig3} show the synthetic profiles of the lines at 458.8, 526.2, and 614.9\,nm analyzed in this study. In these plots, the Stokes $V$ profiles of the FTS atlas were scaled to match the synthesis. Neutral lines, such as \ion{Ca}{i} at 526.2\,nm, are generally better suited than singly ionized lines because the latter disappear in a cool atmosphere such as that present in sunspots and pores. The \ion{Ca}{i} line at 526.2\,nm has small LP (ratio of $V$/LP $\approx 10$) while the lines at 458.8 and 614.9\,nm have zero LP. The only line that has been previously used for calibration purposes is \ion{Fe}{ii} at 614.9\,nm \citep{lites1993,cabrera+bellot+iniesta2005,elmore+etal2010}. 
\section{Application of lines without LP for polarimetric calibration\label{sec:application}} 
The following expression is valid for the Stokes vector $\vec{S}$ after propagation through the telescope optics for a spectral line without LP:
\begin{eqnarray}
\vec{S} = 
\begin{pmatrix}
I\cr
Q\cr
U\cr
V\cr
\end{pmatrix}
= 
\begin{pmatrix}
 T_{11} & T_{12} & T_{13} & T_{14} \cr
    T_{21} & T_{22} & T_{23} & T_{24} \cr
    T_{31} & T_{32} & T_{33} & T_{34} \cr
   T_{41} & T_{42} & T_{43} & T_{44} \cr
\end{pmatrix}
\begin{pmatrix}
I^\prime\cr
0\cr
0\cr
V^\prime\cr
\end{pmatrix}
,\end{eqnarray}
where ${\bf T}$ is the Mueller matrix of the telescope and $(I^\prime, 0, 0, V^\prime)^T$ is the light incident on the telescope.

\begin{table*}
\caption{Transition parameters, line depth and maximal polarization amplitudes at 1500\,G. \label{tab1}\label{tab2}}
\begin{tabular}{cccccccc|c|ccc}
\hline \hline
Element & Ionization & $\lambda_0$            & Excitation & log (gf) & Transition     & g$_{\rm eff}$ & LP? & Line & Max.   & Max.   & Max. \\ 
        & state      &  [nm] & potential (eV)  &          &                & &  &  depth [\%]&  $|Q/I|$  & $|U/I|$  & $|V/I|$                 \\ \hline
Mn      & I          & 425.766          & 2.953      & -0.70    & 4D 0.5- 4P 0.5 & 1.333 & no & 81.0       & 0         & 0         & 0.028  \\ 
Ti      & II         & 431.490          & 1.1609     & -1.104   & 4P 0.5- 4D 0.5 & 1.333 & no  & 92.1       & 0         & 0         & 0.023 \\ 
Cr      & II          & 458.820          & 4.071      & -0.64    & 4D 0.5- 6F 0.5 & 0.333 & no  & 81.2     & 0         & 0         & 0.005    \\ 
Fe      & I          & 514.174          & 2.424      & -2.238   & 3P 1.0- 3D 1.0 & 1.000 & low & 84.3      & 0.016     & 0.007     & 0.092\\ 
Ca      & I          & 526.171          & 2.521      & -0.73    & 3D 1.0- 3P 1.0 & 1.000 & low & 79.7      & 0.012     & 0.004     & 0.085    \\ 
Fe      & II          & 614.923          & 3.889      & -2.8     & 4D 0.5- 4P 0.5 & 1.333 & no & 45.6      & 0         & 0         & 0.019  \\ 
\end{tabular}
\end{table*}

The observed linear and circular polarization are then given by
\begin{eqnarray}
Q &=& T_{21}\cdot I^\prime +  T_{24} \cdot V^\prime\\
U &=& T_{31}\cdot I^\prime +  T_{34} \cdot V^\prime\\
V &=& T_{41}\cdot I^\prime +  T_{44} \cdot V^\prime\;.
\end{eqnarray}
The crosstalk $I\rightarrow QUV \equiv T_{i1} \cdot I$ can easily be measured at continuum wavelengths outside of any spectral line and subsequently be compensated for, which yields:
\begin{eqnarray}
Q_{corr} &=&  T_{24} \cdot V^\prime\label{eqq}\\ 
U_{corr} &=&  T_{34} \cdot V^\prime\label{equ}\\ 
V_{corr} &=&  T_{44} \cdot V^\prime\;.\label{eqv}
\end{eqnarray}
With that, the ratios $Q_{corr}/V_{corr}$ and  $U_{corr}/V_{corr}$ correspond to the ratios of telescope matrix entries $T_{24}/T_{44}$ and $T_{34}/T_{44}$. With a determination of $Q_{corr}/V_{corr}$ and  $U_{corr}/V_{corr}$  from observational data taken at a given telescope, one can thus obtain the ratio of telescope matrix entries at different times and telescope geometries. If a model of the polarimetric properties of the corresponding telescope is available, one can then infer the set of model parameters that best reproduce the observed $Q_{corr}/V_{corr}$ and  $U_{corr}/V_{corr}$ ratios. The approach can also be used in the opposite direction for test purposes. If a telescope correction is applied using a set of parameters and there is some residual LP, it can be deduced that the model parameters are incorrect and need to be updated. For lines with small linear polarization, Eqs.~(\ref{eqq})--(\ref{eqv}) are not strictly valid, but with $V$/LP $\approx 10$ the contribution from $T_{22}, T_{23}, T_{32}$ and $T_{33}$ is still very small.

\begin{figure}
    \centerline{\resizebox{8.8cm}{!}{\includegraphics{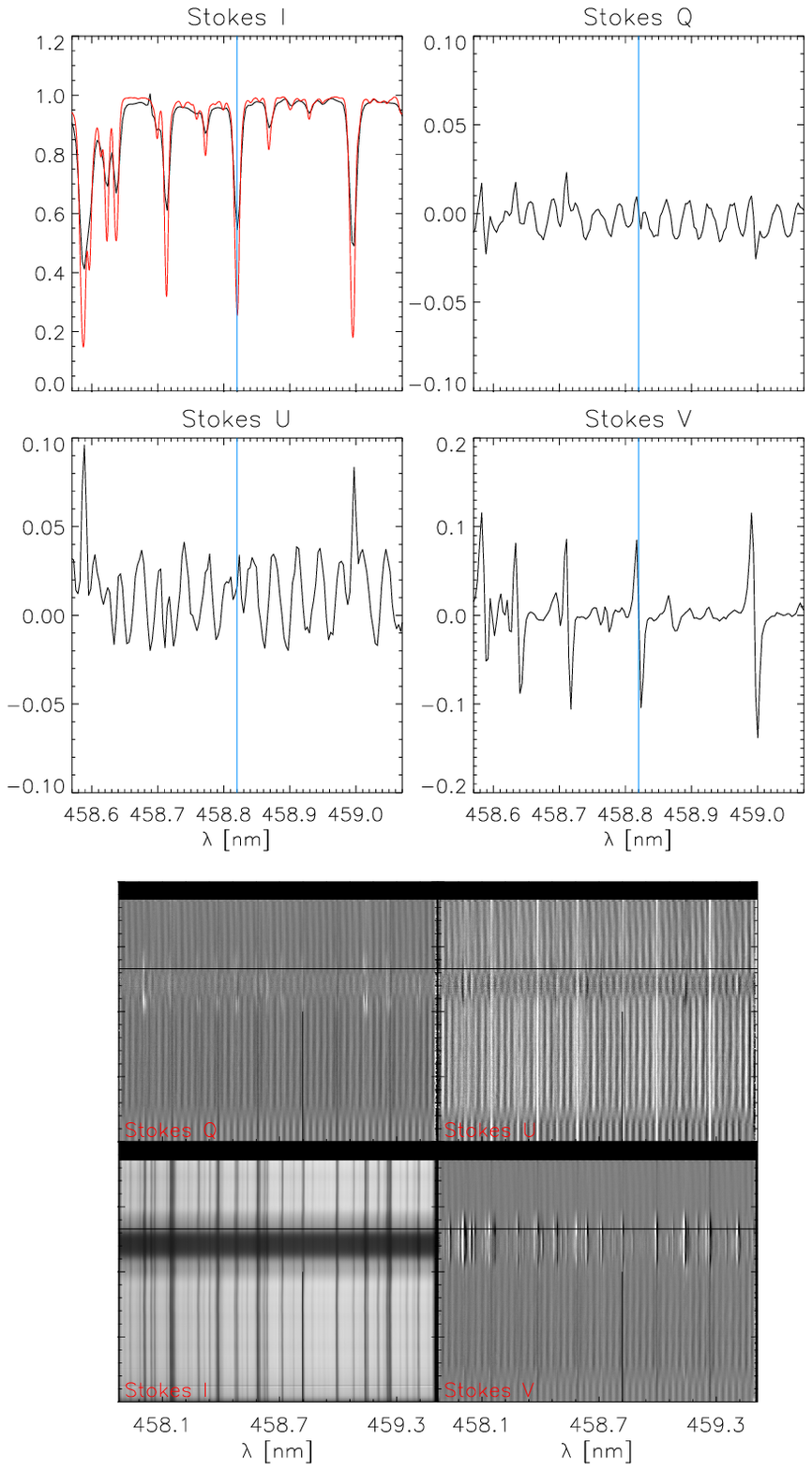}}}
\caption{Observed example spectra at 458.8\,nm. Top panels: individual $IQUV$ profiles from the location indicated with a horizontal black line in the lower panels. The red lines show the corresponding profiles from the FTS atlas. Bottom panels: slit spectra of (clockwise, starting left bottom) $IQUV$ on a cut across the center of a sunspot. The line without LP is indicated by a black vertical bar.\label{fig4}}
\end{figure}

\begin{figure}
  \centerline{\resizebox{8.8cm}{!}{\includegraphics{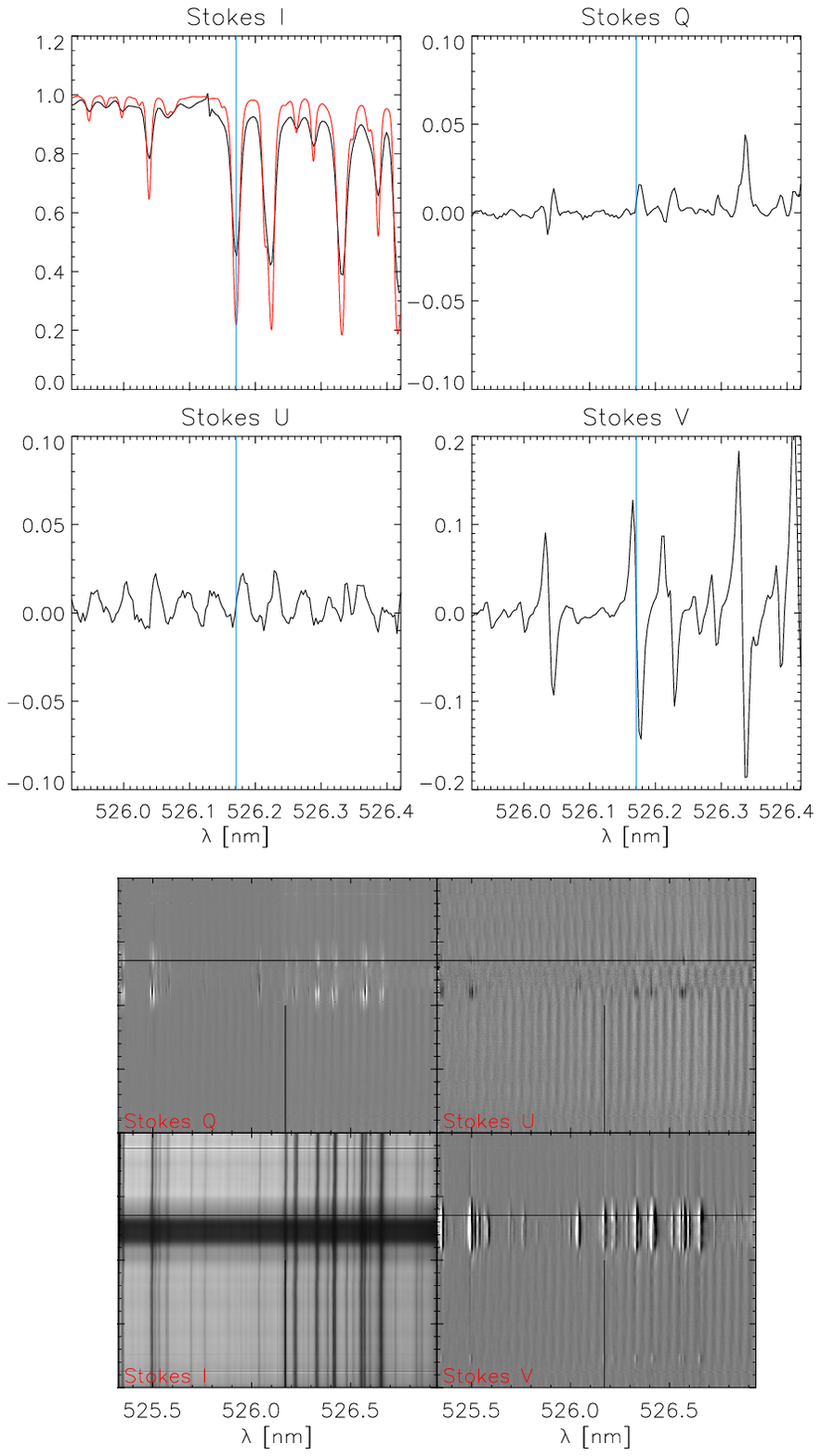}}}
\caption{As in Fig.~\ref{fig4} but for 526.2\,nm.\label{fig5}}
\end{figure}

\begin{figure}
  \centerline{\resizebox{8.8cm}{!}{\includegraphics{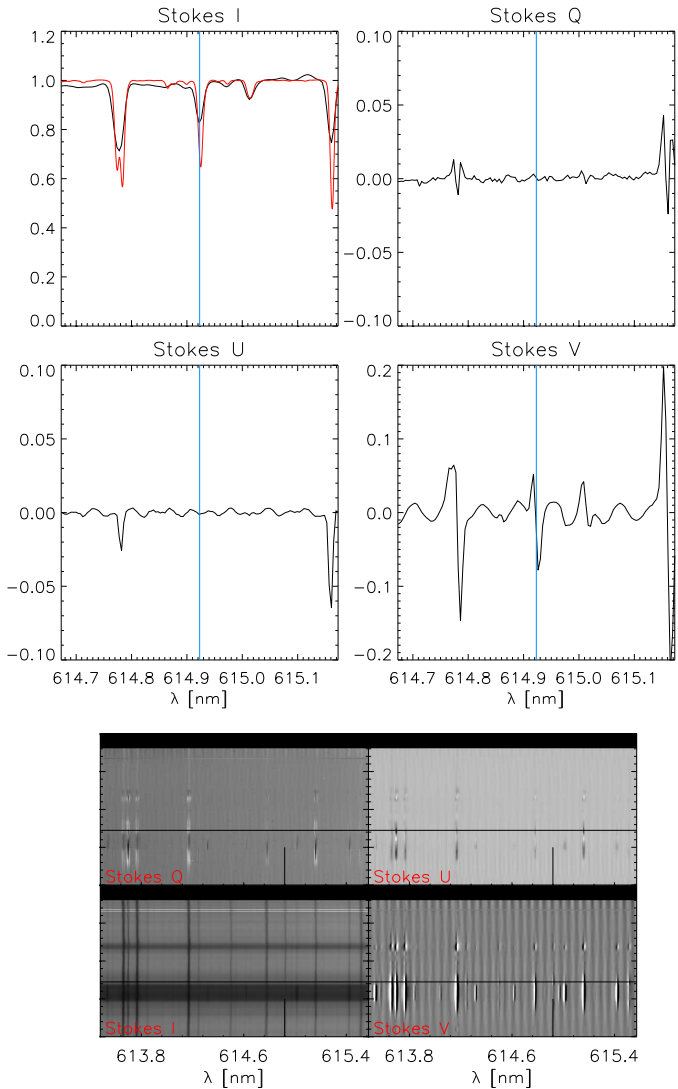}}}
\caption{As in Fig.~\ref{fig4} but for 614.9\,nm.\label{fig6}}

\end{figure}
\section{Observations\label{sec:observations}}
To test the usage of spectral lines without linear polarization for the inference of telescope parameters, we ran an observing campaign at the DST on March 8--31, 2016. We combined the Spectropolarimeter for Infrared and Optical Regions \citep[SPINOR;][]{socasnavarro+etal2006}, the Facility Infrared Spectropolarimeter \citep[FIRS;][]{jaeggli+etal2010} and the Interferometric Bidimensional Spectrometer \citep[IBIS;][]{cavallini2006,reardon+cavallini2008} to obtain spectropolarimetric data in different wavelength regions at the same time. A dichroic beam splitter (BS) was used to separate infrared (IR) and visible (VIS) wavelengths. FIRS was fed with the IR light while the VIS light was split evenly between IBIS and SPINOR by an achromatic 50-50 BS. IBIS and FIRS observed spectral lines with a regular Zeeman pattern (617.3\,nm, 630.25\,nm, 656\,nm, 854\,nm, 1083\,nm, and 1565\,nm) which were not used for the current study. Both IBIS and FIRS were subsequently dropped from the setup on March 17 and 18, respectively, and the corresponding BSs in the light feed were removed or replaced with flat mirrors to increase the light level in SPINOR.

SPINOR was set to observe the six lines without or with small LP listed in Table \ref{tab1} in two different configurations. In the first configuration (setup 1), the lines at 426\,nm, 431\,nm, and 615\,nm were observed from March 8 to 17, while the lines at 459\,nm, 514\,nm, and 526\,nm were covered from March 17 to 24 (setup 2). The wavelength range always covered additional lines with a regular Zeeman pattern as well (Figs.~\ref{fig4}--\ref{fig6}, Figs.~\ref{specexam_426}--\ref{specexam_514}). The spectral sampling was between 3 and 5\,pm. The exposure time was between 22 and 125\,ms and the integration time was between 10 and 30\,s depending on the setup of SPINOR and the light feed. The spatial sampling in the scanning direction was 0\farcs75 (0\farcs37) in setup 1 (setup 2) and the sampling along the slit was about 0\farcs35 in all cases. The slit width was 100$\mu$ corresponding to about 0\farcs75 on the Sun.

In each configuration, several maps of the sunspots in NOAA 12519 and 12524 with $30^{\prime\prime} \times \approx 100^{\prime\prime}$ extent were acquired each day. We also took a few observations of small pores and quiet Sun at disc center to sample regions with smaller polarization amplitudes for comparison. In total, about 30 data sets were obtained in each setup.

Given the amplitude of the interference fringes at all wavelengths shorter than 450\,nm (Figs.~\ref{specexam_426} and \ref{specexam_431}) and the fact that 514\,nm and 526\,nm are close to each other, we only analyzed the data at 459\,nm, 526\,nm, and 615\,nm.

\section{Data analysis\label{sec:data analysis}}
\subsection{Preparation of input data}
This section addresses the methods we used to infer the parameters of the DST telescope model from the observations and the steps required to prepare input data corresponding to Equations (\ref{eqq})--(\ref{eqv}) for all locations with significant polarization signal.
\begin{figure} 
\centerline{\resizebox{8.8cm}{!}{\includegraphics{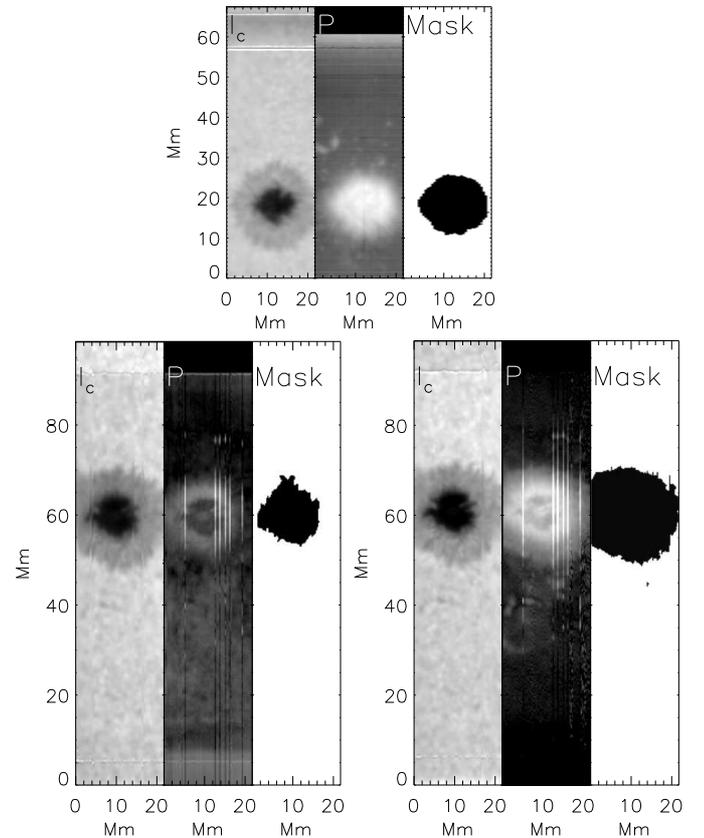}}}
\caption{Maps of one scan at 459\,nm (bottom left panel), 526\,nm (bottom right panel), and 614\,nm (top panel). Left to right: continuum intensity, $I_c$, polarization degree $p_{\rm max}$, and mask of significant polarization signal. The vertical bright stripes in $p_{\rm max}$ for 459\,nm and 526\,nm were caused by a temporary loss of the synchronization between polarization modulation and exposures during the scanning. \label{masks}}
\end{figure}
\subsubsection{Reduction of SPINOR data}
We first reduced the data without applying the correction for the telescope polarization, but with a correction for $I\rightarrow QUV$ crosstalk, using the SPINOR data pipeline\footnote{\url{http://nsosp.nso.edu/dst-pipelines}} that was developed in 2013. The $I\rightarrow QUV$ crosstalk was determined in a continuum wavelength range close to the line of interest without linear polarization. The values of the $I\rightarrow QUV$ crosstalk were stored for later use because they contain the information on the first column of the telescope matrix. We estimate an error of below 1\,\% for the $I\rightarrow QUV$ crosstalk from the presence of the interference fringes (see Appendix \ref{errori2quv}). After this step, a two-dimensional (2D) Fourier filtering in Stokes $QUV$ was applied to reduce the fringe amplitude. The Fourier filter was set to remove spectral frequencies within a manually set frequency range that were constant along the slit, i.e., only fringes with a spatial frequency of zero were removed. 
The fringe pattern in all of the data also contained higher-order fringes that were not taken out. Since the polarization signal can easily extend along 50\,\% of the slit length in observations of active regions, any attempt to include higher spatial frequencies can remove genuine polarization signal as well. The fringe amplitude, period, and phase were also not constant across the spectrum in both the spatial and spectral domain, thus only the primary fringe component could be corrected for. The example slit spectra in Figs. \ref{fig4}--\ref{fig6} show observed spectra after the correction for the strongest fringe pattern. For these example spectra, the telescope correction was applied as well to show the genuine Stokes $QUV$ signals.

\begin{figure}
\centerline{\resizebox{8.cm}{!}{\includegraphics{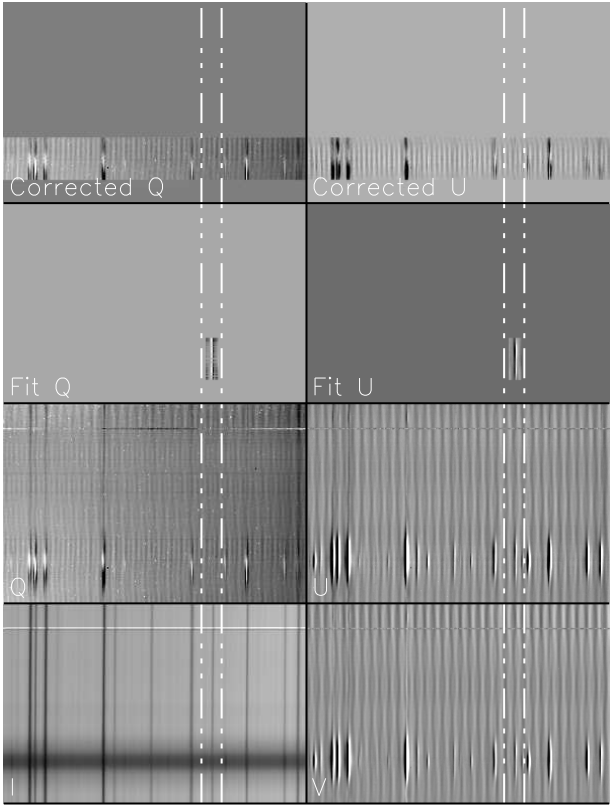}}}
\caption{Example slit spectrum at 615\,nm with the fit of $V\rightarrow QU$. Left column, bottom to top: Stokes $I$, Stokes $Q$, the fitted $Q = \alpha_Q \cdot V$ and the Stokes $Q$ signal after correction, $Q^{\prime} = Q - \alpha_Q \cdot V$. Right column, bottom to top: Stokes $V$, Stokes $U$, the fitted $U = \alpha_U \cdot V$ and the Stokes $U$ signal after correction, $U^{\prime} = U - \alpha_U \cdot V$. The vertical dash-dotted lines enclose the 614.9 nm line. \label{spec_exam1_40}}
\end{figure}
\subsubsection{Determination of locations with significant polarization signal}
To determine the locations inside the FOV that showed significant polarization signal, the maximal polarization degree $p_{\rm max}$ of every profile was calculated by
\begin{eqnarray}
 p_{\rm max} = \mbox{\rm max} [p(\lambda)]_{|\Delta\lambda}= \mbox{\rm max} [\sqrt{Q^2+U^2+V^2}/I(\lambda)]_{|\Delta\lambda}
\end{eqnarray}
in a small wavelength range $\Delta\lambda$ around a spectral line with strong polarization signal. The line need not have been the one with low or zero LP as each spectral range also covered stronger lines (Figures \ref{fig4}--\ref{fig6}).

From the analysis of all profiles, spatial maps of the polarization degree at 459\,nm, 526\,nm, and 614\,nm were created for all observations (see Fig. \ref{masks}). These maps were used to determine the locations of significant polarization signal through a threshold value of 0.01--0.03 by rejecting all profiles with lower polarization amplitudes. By requiring in addition that the profiles above the threshold belong to a spatially connected area with some minimal size, we masked out everything but sunspots and pores. For the 459\,nm data, in several cases, the fringe pattern in the polarization signal was too strong to use the maps of $p_{\max}$. Instead, a threshold in the map of the continuum intensity was used to separate the umbra and penumbra of the sunspot from the brighter quiet Sun and to only retain the profiles located inside the sunspot.
\subsubsection{Determination of $V\rightarrow QU$ crosstalk}
Determination of the $V\rightarrow QU$ crosstalk from the ratio of observed profiles $Q/V$ and $U/V$ can become unreliable for small values of $V$ in the presence of noise. Therefore, a linear fit of $QU_{obs}(\lambda)= \alpha_{QU}V_{obs}$ was made for all profiles above the polarization threshold (see, e.g., Fig. \ref{spec_exam1_40}). Only the close surroundings of the spectral line of interest without LP were used to avoid contamination with line blends.

To verify the quality of the fit, the obtained crosstalk values $\alpha_{QU}$ were used to correct the observed $Q$ and $U$ spectra by $QU^{\prime} = QU - \alpha_{QU} \cdot V$ (top row of Fig.~\ref{spec_exam1_40}). The correction can be applied to the full spectral range. In the example of Fig.~\ref{spec_exam1_40}, the change between Stokes $U$ and $U^\prime$ is obvious, where the latter has no residual LP in the 614.9\,nm line, while all other Zeeman-sensitive lines in the wavelength range have changed from a Stokes $V$-like shape to regular LP signals by the correction. Thus, the fit to first order correctly retrieves the $V \rightarrow QU$ crosstalk values. From the final result, we estimate, however, that the error in $V \rightarrow QU$ caused by the residual fringe pattern is still of the order of a few percent (see Appendix \ref{errorv2qu}).
\begin{figure}
\centerline{\resizebox{8.cm}{!}{\includegraphics{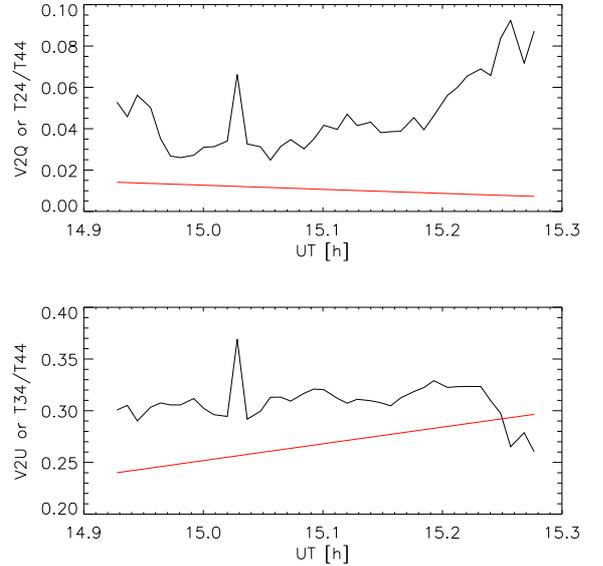}}}
\caption{Crosstalk from $V\rightarrow QU$ averaged along the slit for the map at 614.9\,nm shown in Fig.~\ref{masks} (black lines). The red lines show the values of the DST telescope model for $\frac{T_{24}}{T_{44}}$ and $\frac{T_{34}}{T_{44}}$, respectively, using the telescope parameters determined in 2010. \label{figaqu}}
\end{figure}
\begin{figure*}
\centerline{\resizebox{17.6cm}{!}{\includegraphics{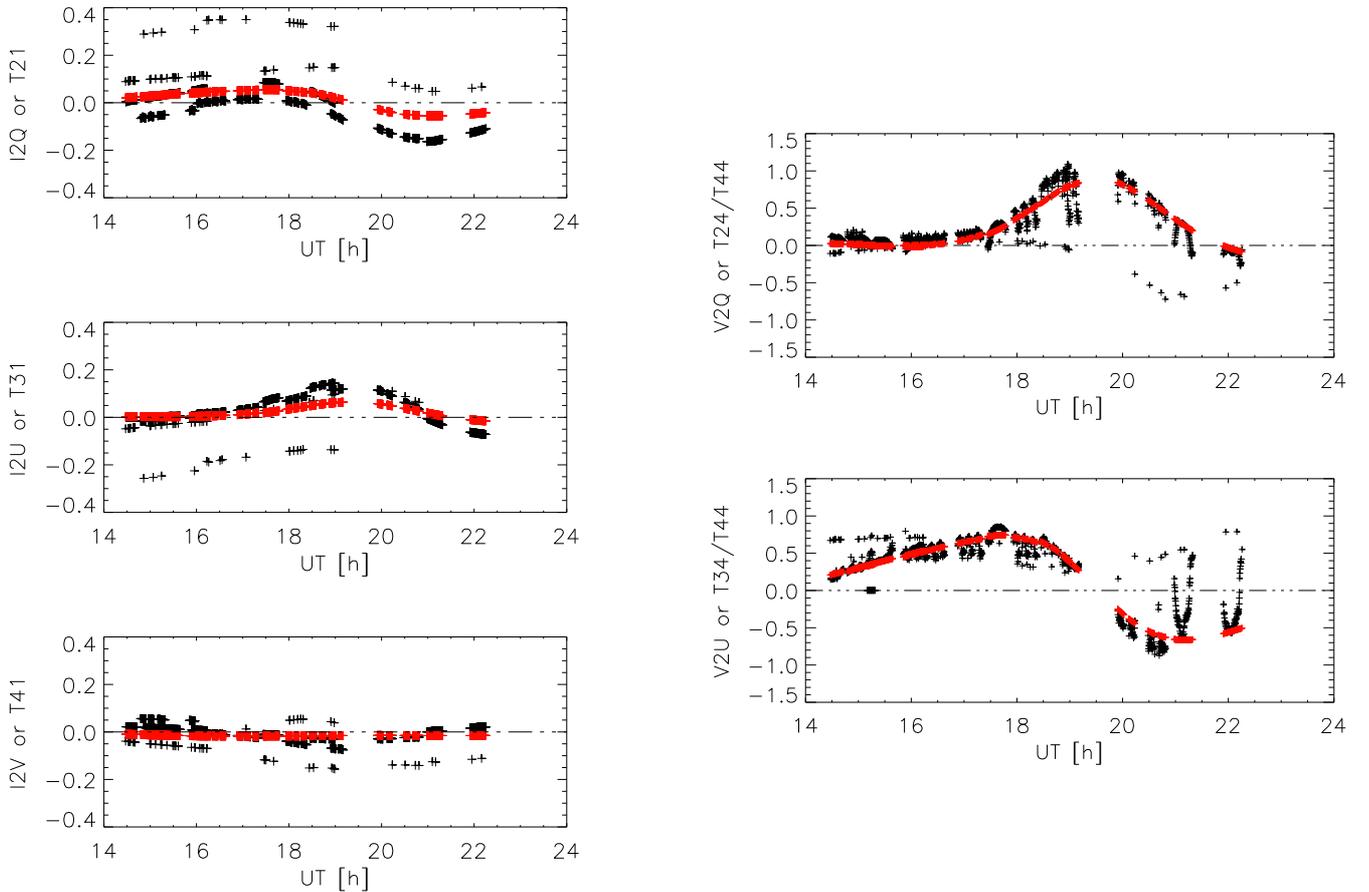}}}
 \caption{Input to the fit of telescope parameters derived from all observations of 526.2\,nm during the campaign. Left column, top to bottom: observed $I\rightarrow QUV$ crosstalk (black crosses). Right column, top to bottom: $V\rightarrow QU$ crosstalk. The red crosses show the corresponding values of the DST telescope model using the 2010 model parameters.\label{parabola}}
\end{figure*}

The values of $\alpha_{QU}$ obtained from the individual profiles were averaged along the slit over all profiles in the mask to retrieve $<\alpha_{QU}>(t)$ because all profiles along the slit were taken at the same time. The time, $t$, is used as a placeholder for the telescope geometry that contains the telescope pointing in elevation and azimuth and the position of the coud{\'e} table (table angle). The numbers used in the calculation of the DST telescope model are actually azimuth($t$), elevation($t$), and table angle($t$). Figure \ref{figaqu} shows the values of $<\alpha_{QU}>(t)$ at 614.9\,nm for one of the observations. For comparison, the values of the corresponding telescope matrix entries predicted by the ``standard'' DST telescope model are overplotted. The set of telescope parameters used was derived from measurements acquired in 2010 which were evaluated using the SCAPA code \citep[see][]{socasnavarro+etal2011}. The modulus of the values in Fig.~\ref{figaqu} agrees to first order, but the agreement is not very close (cf.~Sect.~\ref{fit_select} below).

\subsection{Fit of DST telescope model}
\subsubsection{DST telescope model}
The polarization model of the DST has been described in detail in \citet{skumanich+etal1997} and \citet{socasnavarro+etal2011}. The optical train consists of the entrance window to the evacuated steel tube, the two flat turret mirrors at an angle of incidence (AOI) of 45 degrees in an alt-azimuth mounting, the primary mirror at an AOI of less than 1 degree and the exit window of the vacuum tube. Both turret mirrors are described with the same set of parameters, the ratio of reflectivities parallel and perpendicular to the plane of reflection, $X = r_s / r_p$, and the retardance $\tau_{\rm mirror}$. The primary mirror is modeled as an ideal mirror because of its small AOI. Entrance (EN) and exit (EX) windows are modeled independently as ideal retarders with two parameters each: the retardance  $\tau_{\rm EN, EX}$ and the orientation of the fast axis $\beta_{\rm EN, EX}$. To capture the exact orientation of the telescope model relative to the instrument calibration unit that is located on the rotating coud{\'e} table, a constant offset angle $\theta_{\rm offset}$ between the telescope model and the coud{\'e} table is used as a free parameter. The known geometry of the telescope at any given moment in time is expressed by the elevation, azimuth, and the relative orientation of the coud{\'e} table.

The free parameters of the DST telescope model are usually determined from measurements with a telescope calibration unit (TCU). The TCU consists of an array of sheet polarizers that is placed on top of the entrance window. The TCU is motorized and can be rotated in a full circle. The evaluation of such data and the results are described in detail in \citet{socasnavarro+etal2011}. In the current study, we used the parameter set that was determined using the TCU in 2010 as a reference.

\begin{figure*}
\centerline{\resizebox{17.6cm}{!}{\includegraphics{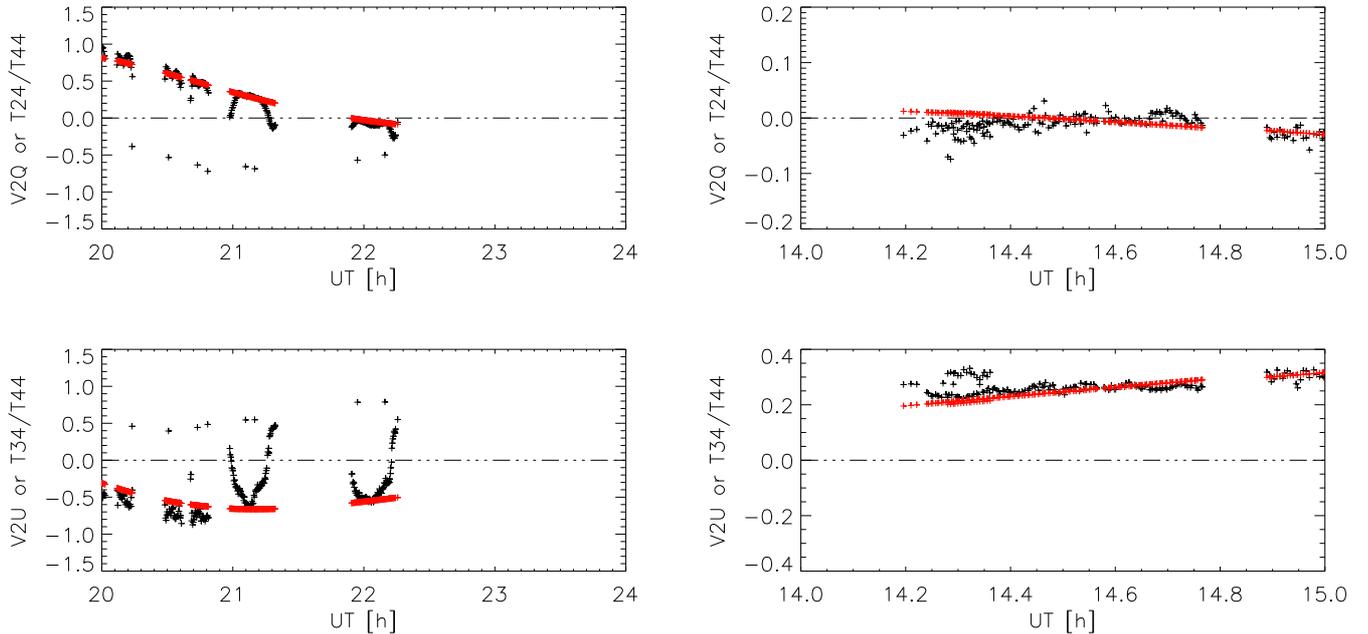}}}
 \caption{Left column: magnification of the $V\rightarrow QU$ crosstalk at 526.2\,nm for observations taken between UT 20 and 24. Individual maps lasted for approximately 30\,min. Most scans show a dominant parabolic shape of the crosstalk values (seen also in Fig.~\ref{parabola}). Only the central position of the scan yields a value that agrees with the 2010 telescope model. Right column: same for data taken in 2017 at 614.9\,nm with the SPINOR modulator in a collimated beam.\label{parabola1}}
\end{figure*}

\subsubsection{Input data for fitting the telescope model\label{fit_select}}
The input data for the fit of the telescope model parameters have two components (see, e.g., Fig~\ref{parabola}), the values of the $I\rightarrow QUV$ crosstalk that correspond to the first column of the telescope matrix, and the values of the $V\rightarrow QU$ crosstalk that correspond to the ratios of the matrix entries $T_{24}/T_{44}$ and $T_{34}/T_{44}$. Both quantities were determined in individual profiles and then averaged along the slit because each slit spectrum corresponds to one moment in time. The geometrical input parameters (elevation, azimuth, table angle) were extracted from the file headers of each spectrum. In the following figures, we plot all the values solely as a function of time for simplicity. The data for each wavelength range were taken within about a week, so the solar position was similar at the same time on different days. All calculations, however, always used the actual telescope geometry at the moment of the observations.

After plotting the corresponding curves for the data at any of the three wavelengths, we noticed a parabolic pattern in the $V\rightarrow QU$ crosstalk data which did not match the 2010 telescope model, but rather touched its predicted values only at the central scan step (see Fig.~\ref{parabola}). The parabolic shape is very obvious for some of the individual maps (see the left half of Fig.~\ref{parabola1}). It presumably was caused by the specific setup for the observations. The SPINOR modulator had been removed from its location in the collimated beam upstream of all instruments. It had been placed as close to the slit as possible to prevent interference with the polarization measurements of FIRS and IBIS and to minimize the image motion caused by a wedge in the modulator optics. The spatial scanning of SPINOR is achieved by moving part of the spectrograph, i.e., the slit unit, the first fold mirror, and the collimator, laterally. The spatial scanning therefore moved the slit relative to the modulator, sampling different areas on the modulator depending on the scan position. The polarimetric calibration of the instrument, however, was only done with the slit centered. Together, this presumably led to a variation of the polarization modulation across the FOV that was not removed by the calibration process. We thus decided to drop all of the data points apart from the central slit position of each map.

We repeated a similar observing run in April 2017 only using SPINOR. In that case, the SPINOR modulator could be left in the collimated beam far upstream of the instrument. The right half of Fig.~\ref{parabola1} shows the corresponding values of the $V\rightarrow Q$ and $V\rightarrow U$ crosstalk for one map observed in 2017 at 614.9\,nm. The parabolic shape is absent, like for all other data taken in 2017, and all of the scan steps with significant polarization signal in this map can be used. This confirms that the behavior in the 2016 data was caused by the mechanical spatial scanning across a spatially resolved modulator close to the slit plane.

\begin{figure*}
\centerline{\resizebox{17.6cm}{!}{\includegraphics{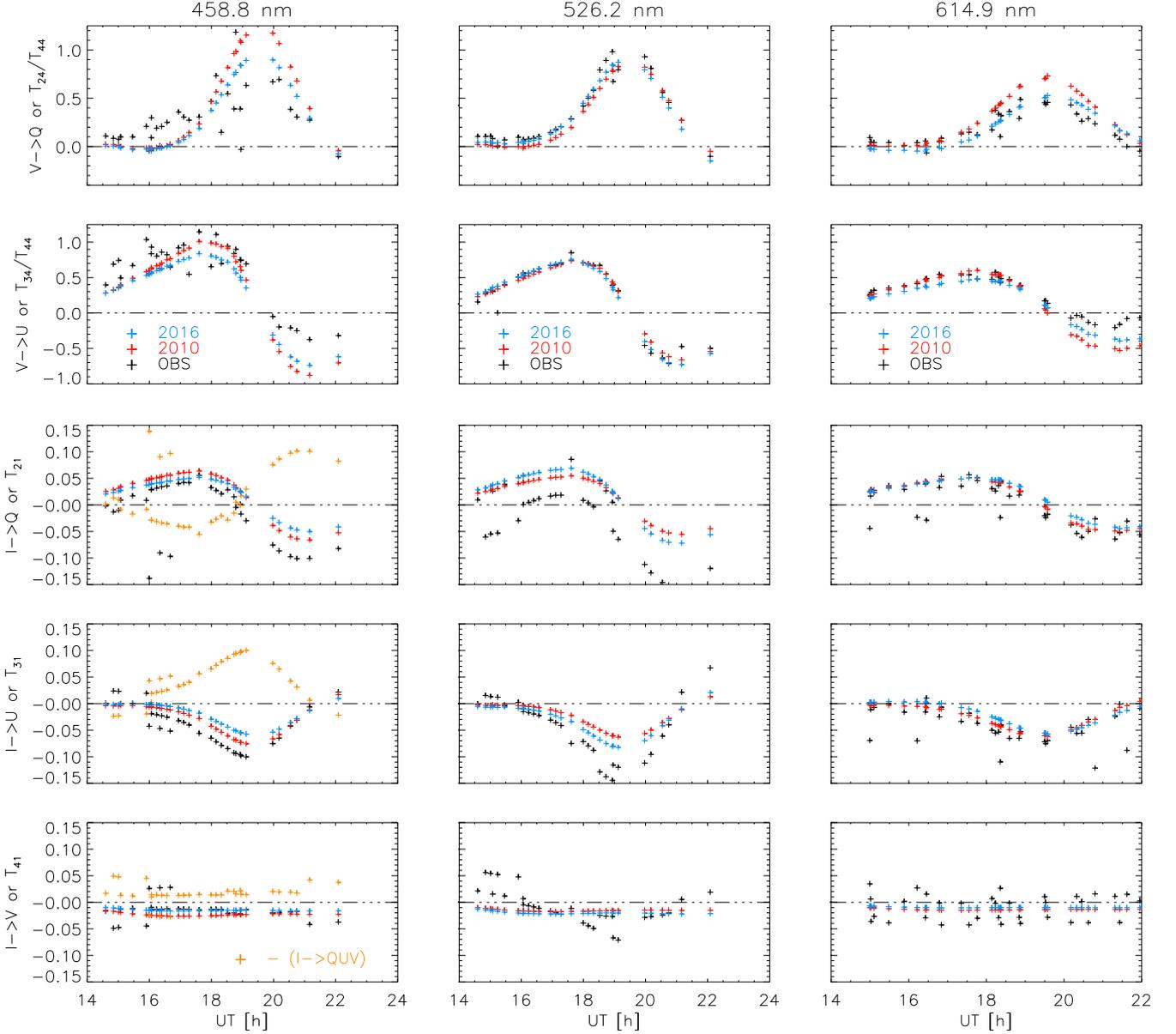}}}
\caption{Fit of telescope parameters at 458.8 nm, 526.2 nm, and 614.9 nm. Top to bottom: observed crosstalk of $V\rightarrow Q$ (black crosses), $V\rightarrow U$, $I\rightarrow Q$, $I\rightarrow U$, and $I\rightarrow V$. The blue (red) crosses show the corresponding values of the DST model using the 2016 (2010) model parameters. The orange pluses in the lower three panels of the leftmost column show the original $I\rightarrow QUV$ values before flipping their sign. \label{sign_flip}}
\end{figure*}
\subsubsection{Fit of telescope model parameters \label{fit_procedure}}
We then attempted to fit the open parameters of the DST telescope model, $X$ and $\tau$ for the turret mirrors, and the retardance  $\tau_{\rm EN, EX}$ and the orientation of the fast axis $\beta_{\rm EN, EX}$ of the entrance and exit windows to the observed $I\rightarrow QUV$ and $V\rightarrow QU$ crosstalk values considering only the central step of each observation. 

It was immediately obvious that there is a sign conflict between the observations and the model in the $I\rightarrow QUV$ crosstalk (orange and red or blue plus symbols in Fig.~\ref{sign_flip}) that were opposite to one another. A mismatch of signs between the 2010 model values determined with SCAPA and the more recent 2013 SPINOR data pipeline was already known. For the application of the telescope correction using the 2010 values in the 2013 pipeline, the offset angle $\theta_{off}$ of about 90 degrees in 2010 had to be changed to $\theta_{off} + 90 \sim$ 180 degrees. There are unfortunately multiple reasons that could cause this sign conflict. A difference in the selection of beams when merging the Stokes vectors would flip the signs of Stokes $Q$, $U$ and $V$. The linear polarizer of the calibration unit (CU) was taken out of and installed back in the CU a few times between 2010 and 2016. If the optical axis got switched by 90 degrees when replacing the polarizer, that also would flip the signs in some of the Stokes parameters.

Given the number of possible options, we were not able to uniquely identify the source for the sign mismatch. We therefore used an ad-hoc correction by flipping the sign of the observed $I\rightarrow QUV$ crosstalk values and used an offset angle of 180 degrees as initial value. The $V \rightarrow QU$ values are not directly affected because they represent a ratio of Stokes parameters. For the evaluation of the data in the current study the sign flip thus has no effect. In the evaluation of solar scientific observations, the sign flip would show up as a change of the azimuth of the magnetic field in the line-of-sight reference frame by 90 degrees.

A second point that became obvious instantly was that the retardances of the windows were not constrained by the data. When trying to fit the retardances of the windows and of the turret mirrors at the same time, the former were running off to unreasonably large values of more than 20 degrees in retardance. Depending on the initial values of the fit, the retardances of the two windows were sometimes changing in opposite directions with approximately a zero net retardance \citep[see also][their Sect.~3]{socasnavarro+etal2011}. 

We therefore modified the fit to a two-step procedure. In the first step, only  $X$, $\tau$ and $\theta_{off}$ were allowed to vary, while in the second step the retardance and position angle of the entrance window and $\theta_{off}$ were varied. The exit window was not considered and set to a unity matrix. The retardance of the mirrors was found to already be able to reproduce the observations very well, so that the window's retardance in the second step stayed close to zero.  
The final fit procedure in our case then only used  $X$, $\tau$, $\theta_{off}$, $\beta_{\rm EN}$ and $\tau_{\rm EN}$.
\begin{table}
\caption{Literature values, best-fit results, and error estimates of telescope parameters.\label{table3}\label{table4}}
\begin{tabular}{ c|c c c c c }
\hline \hline
            & $X$     & $\tau$     & $\theta_{off}$ & $\tau_{\rm EN}$   &$\beta_{\rm EN}$   \\ \hline
          &  \multicolumn{5}{c}{458.8\,nm} \\ \hline
Literature & 1.057 & 165.8 & --           & --             &  --     \\   
2010 TCU      & 1.101 & 148.9 & 93.1       & 3.0        & 138.6   \\ 
2016 fit        & 1.077 & 153.2 & 178.2      & 0.6          & 22.0     \\ 
$\sigma$ &      $\pm$ 0.033  &  $\pm$ 0.5  &  $\pm$ 0.7 &   $\pm$ 0.6 & -- \\ \hline          
           & \multicolumn{5}{c}{526.2\,nm} \\ \hline
 Literature & 1.059 & 167.7 & --           & --            &  --        \\
 2010 TCU      & 1.083 & 155.0 & 93.1       & 2.3        & 138.6   \\ 
 2016 fit        & 1.108 & 154.3 & 182.0      & 0.6         & 22.0     \\ 
$\sigma$ & $\pm$ 0.035  &  $\pm$ 0.5 &  $\pm$ 0.7 &   $\pm$ 0.6 & -- \\ \hline
           &  \multicolumn{5}{c}{614.9\,nm} \\ \hline
 Literature & 1.065 & 169.5 & --           &  --           &   --      \\
 2010 TCU      & 1.080 & 157.8 & 93.1       & 2.1        & 138.6  \\ 
 2016 fit        & 1.074 & 162.7 & 174.2      & 0.4         & 22.0   \\ 
$\sigma$ & $\pm$ 0.036  &  $\pm$ 0.6  &  $\pm$ 1.1 &   $\pm$ 0.8& -- \\ 
  $\chi^2 \pm 10 \%$ &  $\pm$ 0.150  &  $\pm$ 2.6  &  $\pm$ 4.0 &   $\pm$ 4.0 & -- \\\hline
\end{tabular}
\end{table}
\section{Results \label{sec:results}}
\subsection{Fit quality}
Figure \ref{sign_flip} shows the final result of the fit (blue pluses) together with the corresponding curves when using the 2010 model parameters (red pluses) and the observations (black pluses) for all three wavelengths. The input data at 458.8\,nm show the largest scatter that presumably results from the residual fringe pattern (Fig.~\ref{fig4}). The differences between the new determination and the 2010 values are rather small, only a few percent. Both parameter sets provide a good fit to the input data. The match of observations and telescope model with any parameter set is slightly worse in the afternoon for times later than about UT 19:00.

\subsection{Ratio of reflectivities $X =  r_s/r_p$ \label{sect_x}}
Table \ref{table3} lists the results for all free fit parameters from the current determination and the 2010 measurements. In addition, we calculated the values of $X$ and $\tau$ for the turret mirrors assuming optically thick coatings (i.e., no influence of the underlying glass substrate), using the literature values at each wavelength given in the Handbook of Chemistry and Physics \citep{lide1994}. For the ratio of reflectivities $X$, there is no clear wavelength trend in either the 2010 or 2016 results, while the literature values show a monotonic decrease with increasing wavelength. All values for $X$ determined from measurements at the DST are around 1.09$\pm$0.1, while the literature values are about 0.02-0.05 lower \citep[see also][their Fig.~8]{socasnavarro+etal2011}. The value of $X$ is only weakly constrained by the  $V\rightarrow QU$ crosstalk. It cancels out for a single mirror \citep[see, e.g.,][]{beck+etal2005a} because, for instance,
\begin{equation}
V\rightarrow Q \equiv T_{24}/T_{44} = \frac{-2\,X\,\sin\,\tau}{2\,X\,\cos\,\tau} = \tan\,\tau \neq \mbox{f}(X)\,. \label{eq9}
\end{equation}
At the DST, there is a slight dependence of $V\rightarrow Q$ on X because the two turret mirrors rotate relative to each other. The sensitivity test described in Sect.~\ref{sensitivity} below confirms this. Within our assumed error range, the measurements of $X$ from all data acquired at the DST agree well.
\begin{figure}
\centerline{\resizebox{8.8cm}{!}{\includegraphics{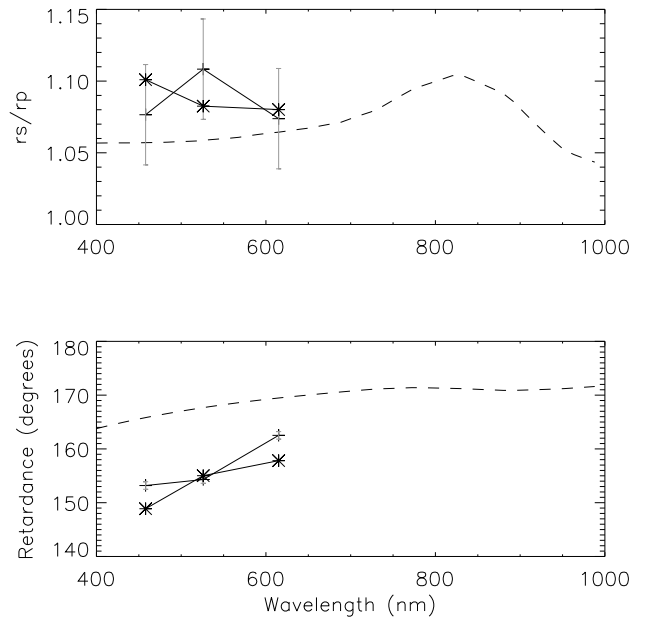}}}
\caption{Values of $X$ (top panel) and $\tau$ (bottom panel) for the literature values (dashed lines), the 2010 telescope model parameters (asterisks), and the current 2016 fit results (crosses). The vertical gray lines show error estimates of $\pm$ 1 degree (bottom panel) and $\pm$ 0.035 (top panel). \label{rsrpandret}}
\end{figure} 
\subsection{Window and mirror retardance $\tau$}
The mirror retardance alone was found to already cover the needed values and to provide a good fit to the amplitude of the daily variation in $V\rightarrow QU$. The second step of the two-step fit using the entrance window retardance as described in  Sect.~\ref{fit_procedure} above yielded a window retardance $\tau_{\rm EN}$ of less than 1 degree, therefore making the window seem negligible. We note that the initial value of the orientation of the optical axis of the window was 22 degrees. This was not significantly changed in the fit (last column of Table \ref{table3}), which would support the assumption that the window does not contribute to the polarization properties of the optical train, or its contribution cannot be separated from that of the mirrors given the input data. 

The retardance of the turret mirrors $\tau$ increases monotonically with wavelength in all cases, i.e., for the 2010 fit, the 2016 fit, and the literature values. The 2010 and 2016 fit have values that differ by about 5 degrees, while the literature values are offset from them by more than 10 degrees. The literature values do not take any information obtained at the DST into account, so their global offset from both fit values is uncritical, whereas the fit values themselves agree to first order.

Figure \ref{rsrpandret} shows a plot of both $X$ and $\tau$ for the turret mirrors for all available values for completeness. The error bars were taken from the estimates in the following section.

\begin{figure*}
\sidecaption
  \centerline{\resizebox{15.6cm}{!}{\includegraphics{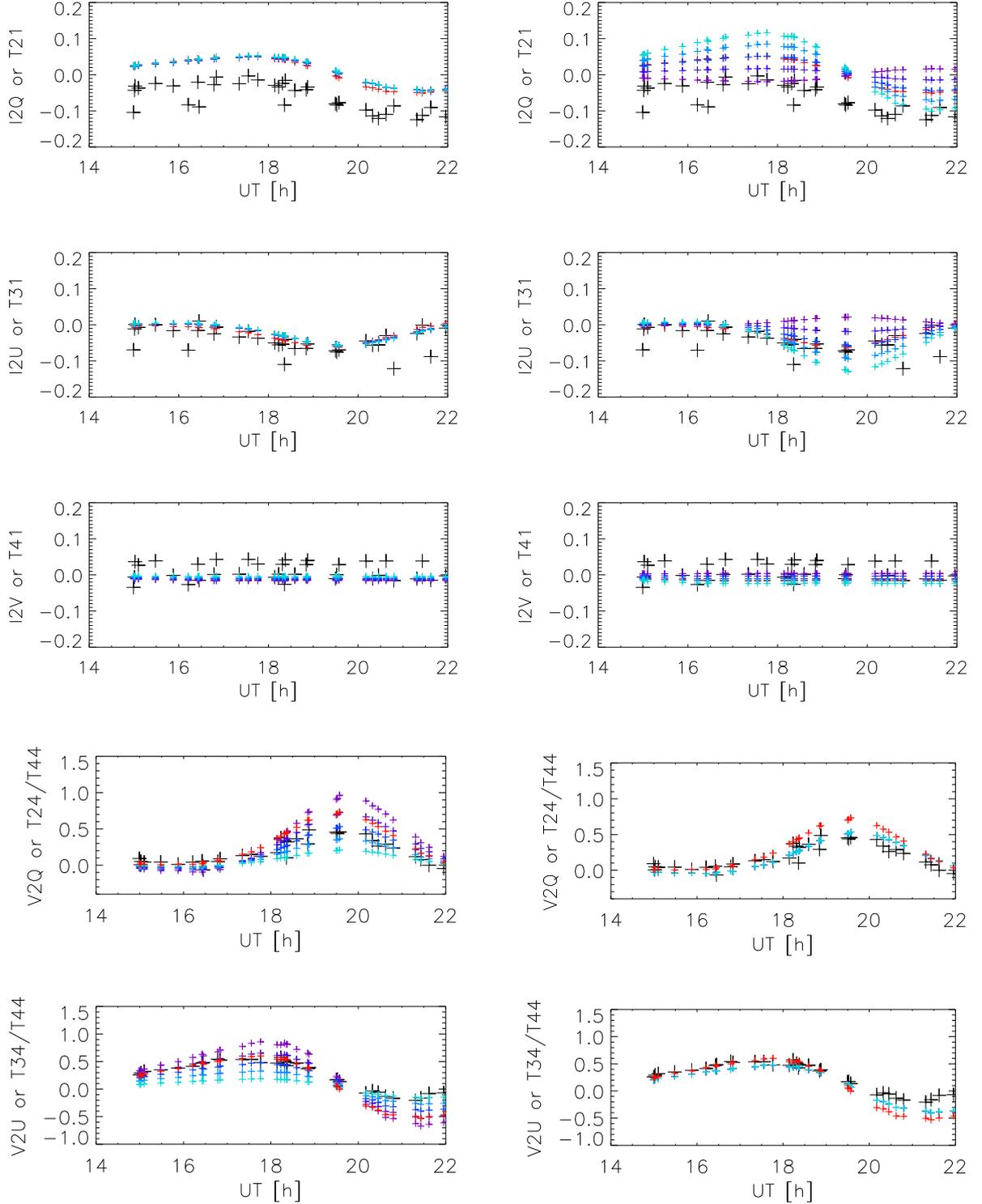}}}
\caption{Sensitivity test of the telescope model at a wavelength of 614.9\,nm. Top three rows: change in $I\rightarrow QUV$ for a variation of the mirror retardance $\tau$ (left column) and the ratio of reflectivities $X$ (right column). Observations are given by black plus symbols, while the red plus symbols correspond to the 2010 parameter set. $\tau$ and $X$ were varied from -10 to 10 and -0.1 to 0.1, respectively. The resulting curves are indicated by the series of blue to purple plus symbols. Bottom two rows: change in $V\rightarrow QU$ for a variation of the mirror retardance $\tau$ (left column) and the ratio of reflectivities $X$ (right column).\label{sensi_fig}}
\end{figure*}
\subsection{Sensitivity test and error estimate \label{sensitivity}}
To check the sensitivity of the fit and the telescope model to changes in the model parameters, we varied $X$ and $\tau$ of the turret mirrors by -0.1 to 0.1 and -10 to 10 around the best-fit value, while keeping the window properties at the best-fit values. Figure \ref{sensi_fig} shows the reaction of the telescope model in the $I\rightarrow QUV$ and $V\rightarrow QU$ crosstalk values to these changes at a wavelength of 614.9\,nm. The other two wavelengths at 459\,nm and 526\,nm showed a similar behavior. 
 
The $V\rightarrow QU$ crosstalk reacts only weakly to changes in $X$. Only for values of $X$ that were far outside the range of variation used here, e.g., $X <0.5$ or $X>2$, was a significant change seen. This is due to the fact that the two turret mirrors rotate relative to each other when tracking the Sun, while $X$ cancels out for a single mirror as shown by Eq.~(\ref{eq9}) above. The $I\rightarrow QUV$ crosstalk on the other hand reacts strongly to changes in $X$, while it is nearly unaffected by changes in $\tau$. 

We used the sensitivity test to estimate a reliability interval for the fit procedure by determining the variation of $X$ or $\tau$ that led to a change of the $I\rightarrow QUV$ or $V\rightarrow QU$ crosstalk by 0.01 on average. The scatter in the observational values, e.g., for $I\rightarrow QUV$ (see Appendix \ref{errestim}), is of about that order of magnitude, and a variation of 1\,\% on average caused a clearly visible displacement between the observations and the model output. For instance, for determining the error of $X$, we calculated the average difference between the observed and telescope model values by
\begin{eqnarray}
D = \frac{1}{3\,N} \, \sum_{i=1...N} \sum_{j=2...4} | T_{j1}^{obs} - T_{j1} (X,\tau)|_i \,,
\end{eqnarray}
where $N$ is the number of data points and $T_{j1}^{obs} = I\rightarrow Q,U,V$ for $j=2,3,4$. 

We then modified the best-fit solution $X_{best-fit}$ by  $\pm \Delta X$ until a value of $D=0.01$ was reached. For the error of $\tau$, we used  $V\rightarrow QU$ in the calculation instead. This yielded a range of about $\pm 0.035$ for $X$, $\pm (0.5-0.8)$ degrees for the retardances $\tau$ of the turret mirrors and the entrance window and $\pm 1 $ degree for the offset angle (see Table \ref{table3}). At the very small retardance best-fit value of the entrance window of less than a degree, the orientation of the window's fast axis was nearly unconstrained. Neither $\chi^2$ nor the curves from the telescope model over the day showed a significant response to changes in $\beta_{EN}$. We therefore could not derive any reasonable error estimate for this parameter. All three spectral ranges gave very similar results. As a second estimate for the error, we varied the parameters around the best-fit solution at 614.9\,nm until $\chi^2$ changed by $\pm 10$\,\%. If the $\chi^2$-surface near the best-fit solution can be approximated by an $n$-dimensional Gaussian, where $n$ is the number of free parameters, then a change of $\chi^2$ by 10\,\% corresponds to a width of the Gaussian by about one $\sigma$. This yielded errors that were larger by a factor of about 4, which seems to be somewhat of an overestimate of the true error.
\begin{table}
\caption{Fit parameters at 614.9\,nm for three different fixed values of the offset angle.\label{tab_theta}}
\centering
\begin{tabular}{cccc}
\hline\hline
$\chi^2$ & $X$ & $\tau$ & $\theta_{off}$ \cr\hline
 6.167 & 1.078   &    162.4 &    182.0\cr
 5.192 & 1.075   &    162.0 &   178.1\cr
 4.808  &1.071    &   162.7  &   174.4\cr
\end{tabular}
\end{table}
\begin{figure*}
\centerline{\resizebox{17.6cm}{!}{\includegraphics{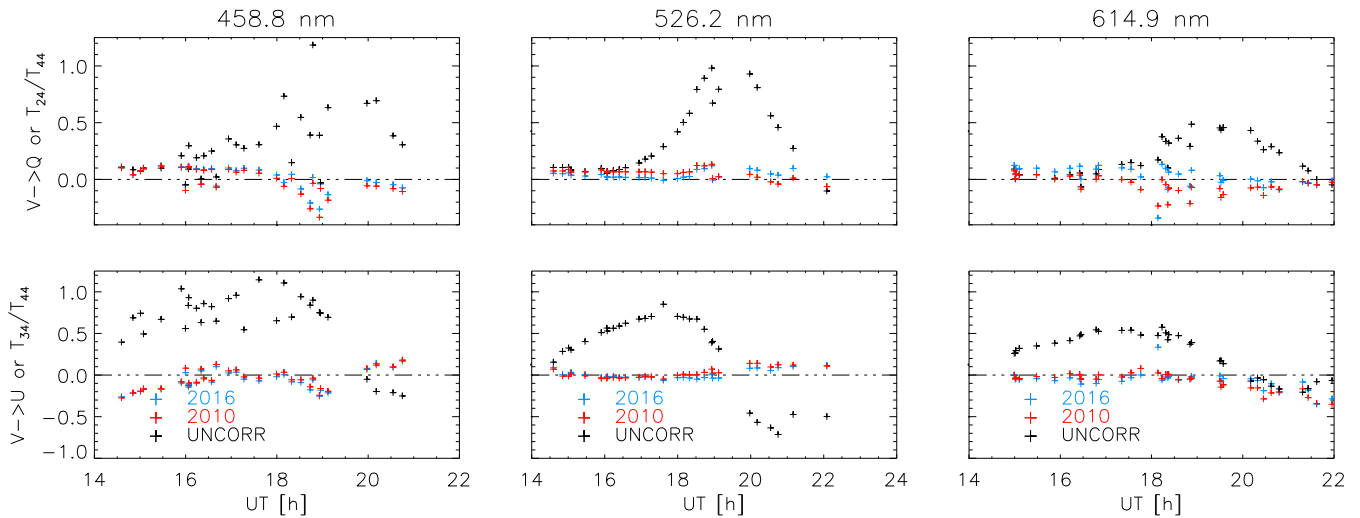}}}
\caption{Residual $V\rightarrow Q$ (top row) and $V\rightarrow U$ (bottom row) crosstalk after applying the telescope correction. Left to right: at 458.8\,nm, 526.2\,nm, and 614.9\,nm. Black plus symbols: observations without telescope correction. Red/blue plus symbols: observations with a telescope correction using the 2010/2016 model parameters.\label{residual_ct}}
\end{figure*}

The offset angle varies by up to 8 degrees between the different wavelengths even if it should reflect rather a mechanical, constant property than any wavelength-dependent optical property. To test the impact of this large variation on the other parameters, we re-ran the fit at 614.9\,nm for the maximal, minimal, and average value of $\theta_{off}$ while keeping the offset angle fixed. Table \ref{tab_theta} shows that the value of  $\theta_{off}$ has an impact on the final $\chi^2$ of the fit, but only leads to a minor variation in the mirror parameters $X$ and $\tau$ that is within their estimated error range. The offset angle represents a physical rotation and changes the phase of the daily variation and the relative fraction of Stokes Q and U, whereas the main constraint for the mirror parameters is the amplitude of the daily variation (Fig.~\ref{sensi_fig}). 

\subsection{Residual crosstalk with telescope correction}
To test the fit quality, we applied the telescope correction to the observations using the parameter set from 2010 and 2016 prior to running the determination of the $I\rightarrow QUV$ and $V\rightarrow QU$ crosstalk. With the telescope correction, all crosstalk values should be close to zero. Figure \ref{residual_ct} shows that this is indeed the case for the $V\rightarrow QU$ crosstalk. The rms fluctuations of the $V\rightarrow QU$ crosstalk are given in Table \ref{table5} for all wavelengths. With the application of the telescope correction, they reduce from the uncorrected case with about 20--30\,\% rms to 3--10\,\% rms after the correction.  The amplitude of the residual crosstalk in fully calibrated data after the application of the telescope correction matches the fit residuals (see Appendix \ref{errorv2qu}) which correspond to the mismatch between observations and model prior to the application of the inverse telescope matrix. The new determination in 2016 yields a slightly lower rms than the 2010 parameter set, but usually only by about 1\,\%.  The residual crosstalk level thus remains at a value of 3-10$\times10^{-2}$ regardless of which correction is applied. 

\begin{table}
\caption{Residual rms $V\rightarrow QU$ crosstalk. \label{table5}}
\centering
\begin{tabular}{ c|c c }
\hline \hline
                & $V\rightarrow Q$   & $V\rightarrow U$   \\ \hline
                &\multicolumn{2}{c}{458.8\,nm} \\ \hline
 Uncorrected    & 0.27 & 0.43   \\ 
 2010 TCU & 0.11 & 0.12  \\ 
 2016 fit            & 0.10 & 0.12  \\ \hline
                & \multicolumn{2}{c}{526.2\,nm} \\ \hline
 Uncorrected    & 0.33 & 0.47   \\ 
 2010 TCU & 0.04 & 0.06  \\ 
 2016 fit            & 0.03 & 0.05  \\ \hline
                &  \multicolumn{2}{c}{614.9\,nm} \\ \hline
Uncorrected    & 0.17 & 0.25    \\ 
 2010 TCU & 0.08 & 0.11   \\ 
 2016 fit            & 0.09 & 0.11   \\                  

\end{tabular}
\end{table}

\section{Summary \& Discussion \label{sec:discussion}}
We acquired a series of spectropolarimetric observations of active regions containing sunspots over a few days in six different spectral lines without or with negligible linear polarization signals (cf.~Table~\ref{tab1}). From the analysis of the crosstalk between intensity and polarization, $I\rightarrow QUV$, and the crosstalk between circular and linear polarization, $V\rightarrow Q$ and $V\rightarrow U$, we were able to infer the parameters that characterize the model of the polarization properties of the DST for a subset of three spectral lines, and hence three wavelength regions at 459, 526, and 615\,nm (cf.~Table~\ref{table3}). 
\subsection{Method}
Unlike most other calibration techniques in use for current ground-based solar telescopes \citep{skumanich+etal1997,beck+etal2005a,selbing2010,socasnavarro+etal2011}, this approach does not require any additional optics other than a spectropolarimeter with medium to high spectral resolution. The method does not rely on any assumptions for lines without intrinsic linear polarization, while for the other lines the linear polarization is assumed to be negligibly small  \citep{lites1993,VillahozandAlmeida+1993,VillahozandAlmeida+1994,li+etal2017}.

The analysis procedure is automatic to a high degree. After identification of the corresponding spectral line to be used, the thresholds in polarization degree or continuum intensity for the selection of spatial positions to be included and the initial values for the least-squares fit are the only items which must be manually provided.

The method is based on a statistical approach. The data to be analyzed do not have to be acquired consecutively over a few days as done in our case, but can be collected over a longer period of time as long as the telescope properties do not vary significantly within that time frame. In addition, the spatial resolution of the observations does not need to be very high because the generation of crosstalk between polarization states only requires a significant signal in the source polarization state. This offers the possibility to build up a base data set with sufficient statistics over a time frame of weeks to months during periods of bad seeing. 

In the same way that high spatial resolution is not ultimately required, there is also no requirement for observing active regions, pores, or sunspots. Polarization amplitudes of a few percent in Stokes $V$ are also reached by magnetic network elements \citep[e.g.,][]{rezaei+etal2007}. An accurate determination of the crosstalk between polarization states then only requires a high signal-to-noise ratio to detect signals of a fraction of a percent, i.e., signals of 10$^{-4} - 10^{-3}$ amplitude must be above the noise floor.

Several of the spectral lines without, or with negligible linear polarization signal that can be used for the application of the method (see Tables \ref{table_wolp} and \ref{table_wslp}) result from neutral atoms, unlike the singly ionized line at 614.9\,nm. The advantage of lines pertaining to neutral elements is that their line strength does not weaken in ``cold'' solar structures such as pores and sunspots, while singly ionized lines can disappear completely exactly where the magnetic field strength and polarization signal is largest (see Fig.~\ref{fig6}). 

Although there is a dense wavelength coverage from the blue end of the visible spectrum at 400\,nm up to 615\,nm, there were no suitable lines found in the red end of the spectrum ($>$\,615\,nm).

\subsection{Performance}
As for most methods used to determine polarization properties, it is difficult to provide a good estimate of the accuracy of the approach. We find a residual $V\rightarrow QU$ crosstalk of about 5-10\,\% after application of the correction for the telescope polarization based on our 2016 best-fit values (Table \ref{table5}). The residual crosstalk is comparable to that when using the 2010 telescope parameter set that was derived with a TCU. However, as seen in Figs.~\ref{sign_flip} and \ref{residual_ct}, there is already a significant scatter of a few percent in the input data that is presumably caused by residual interference fringes, and is thus due to the quality of the input data, not the approach itself. 

The values of the telescope parameters of our current determination match those of the 2010 telescope model within the error bars, however both are offset from literature values (Fig.~\ref{rsrpandret}). The match of the values determined from actual measurements at the DST is a positive sign, as the applicability of the literature values assuming thick coatings to the DST mirrors is not ensured. Therefore, based on our findings, we propose that this new calibration method can be just as accurate as the standard DST calibration with a TCU depending on the data quality. 

The main limitation for the application of the approach to the current data is the interference fringes in the observations; they impact the determination of crosstalk values between different Stokes parameters at a level of several percent.

\section{Conclusions \& future work \label{sec:conclusions}}
We conclude that it is possible to derive the parameters that describe the polarization properties of a telescope from observations of spectral lines without, or with negligible linear polarization signal. These spectral lines cover much of the blue side of the visible part of the spectrum, but no suitable lines were found above 615\,nm.  

It is impossible to use a conventional telescope calibration unit consisting of linear polarizers for the 4-m DKIST telescope. Using spectral lines without intrinsic linear polarization is thus a promising approach for its polarization calibration. At the DST, a time-dependent system with a variable geometric configuration has to be calibrated. The corresponding problem for DKIST is much less complex as only the first two mirrors need to be characterized which are static with fixed angles of incidence and a fixed relative orientation. 

We would also like to search for suitable lines towards the red end of the spectrum and develop a similar method for lines with a regular Zeeman pattern, for which one would need to make assumptions either about the magnetic field geometry or about the symmetry properties of Zeeman signals.

\begin{acknowledgements}
The Dunn Solar Telescope in Sunspot/NM is operated by the National Solar Observatory (NSO). The NSO is operated by the Association of Universities for Research in Astronomy, Inc. (AURA) under cooperative agreement with the National Science Foundation.
\end{acknowledgements}
\bibliographystyle{aa}
\bibliography{references_luis_mod}

\begin{appendix}
\section{Line spectra at 426, 431, and 514\,nm\label{appa}}
\begin{figure*}
 \centerline{\resizebox{17.6cm}{!}{\includegraphics{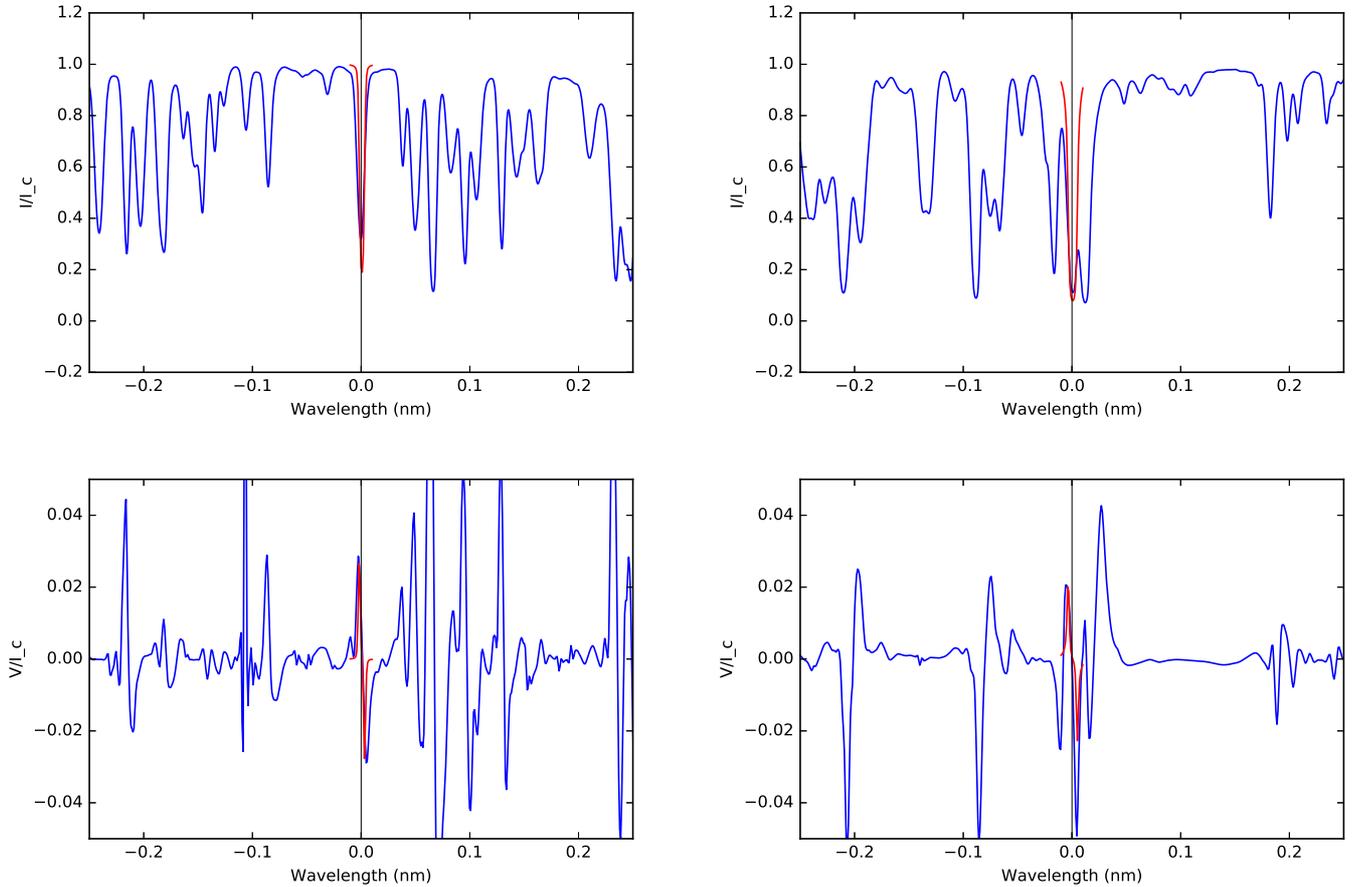}}}
  \caption{Synthetic Stokes profiles of Stokes $I$ (top panel) and $V$ (bottom panel) for the \ion{Mn}{i} line at 425.766\,nm (left column) and \ion{Ti}{ii} at 431.490\,nm in the synthesis with 1500\,G. The blue lines show the corresponding profiles from the FTS atlas.\label{fig_synth5}}
\end{figure*}
\begin{figure*}
  \centerline{\resizebox{17.6cm}{!}{\includegraphics{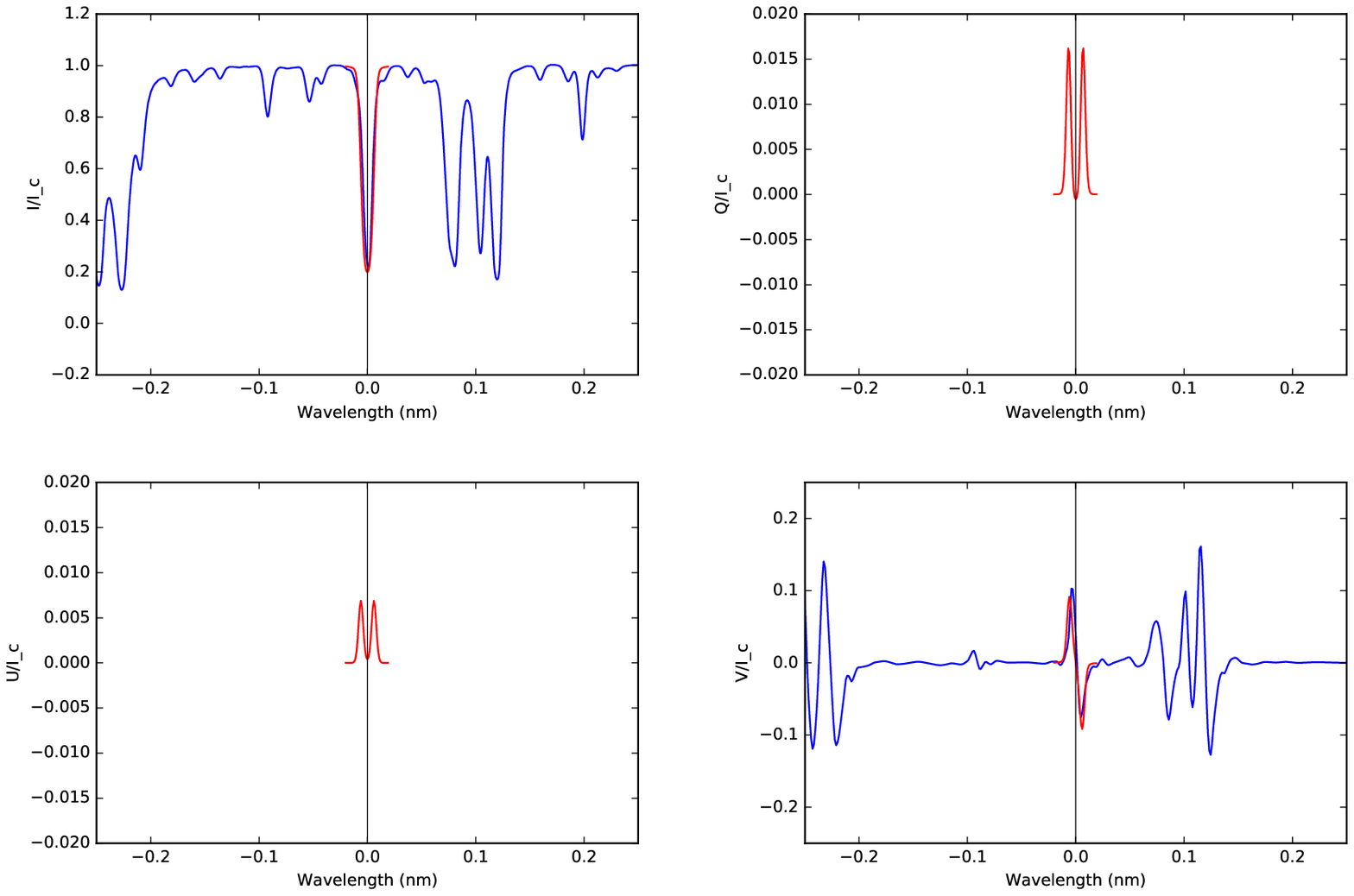}}}
  \caption{Synthetic Stokes profiles of Stokes $I$ (top left), $Q$ (top right), $U$ (bottom left) and $V$ (bottom right) for the \ion{Fe}{i} line at 514.174\,nm in the synthesis with 1500\,G.\label{fig_synth6}}
 \end{figure*}
Figures \ref{fig_synth5} and \ref{fig_synth6} show the profiles of the lines of \ion{Mn}{i} line at 425.766\,nm (no LP), \ion{Ti}{ii} at 431.490\,nm (no LP), and \ion{Fe}{i} line at 514.174\,nm (small LP) from the synthesis with a magnetic field strength of 1500\,G.

\begin{figure*}
\centerline{\resizebox{17.6cm}{!}{\includegraphics{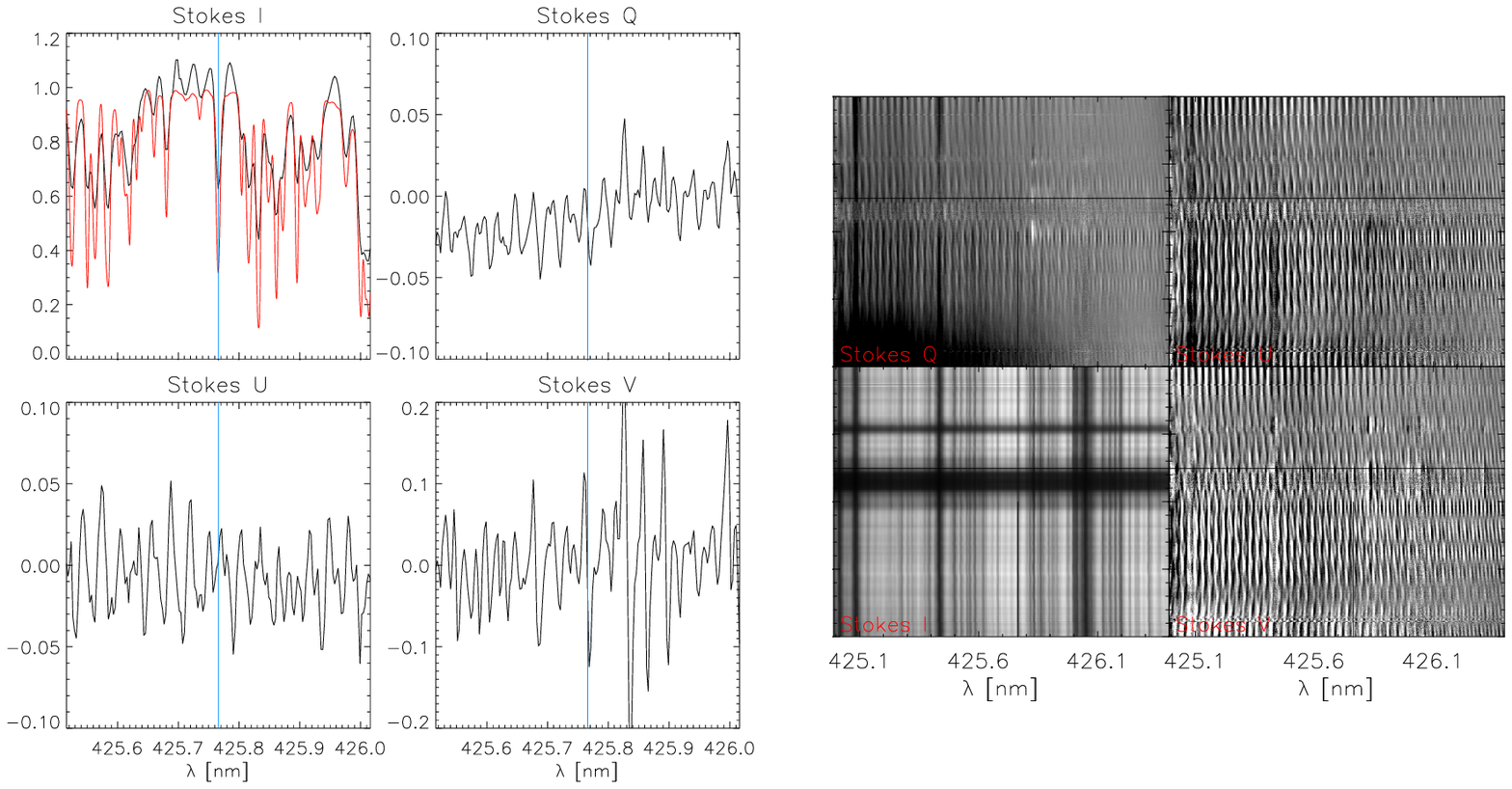}}}
\caption{Observed example spectra of \ion{Mn}{I} at 425.8\,nm. Left four panels: individual $IQUV$ profiles from the location indicated with a horizontal black line in the right panels. The red line shows the corresponding profile from the FTS atlas. Right four panels: slit spectra of (clockwise, starting left bottom) $IQUV$ on a cut across the center of a sunspot. The line without LP is indicated by a black vertical bar.\label{specexam_426}}
\end{figure*}
\begin{figure*}
\centerline{\resizebox{17.6cm}{!}{\includegraphics{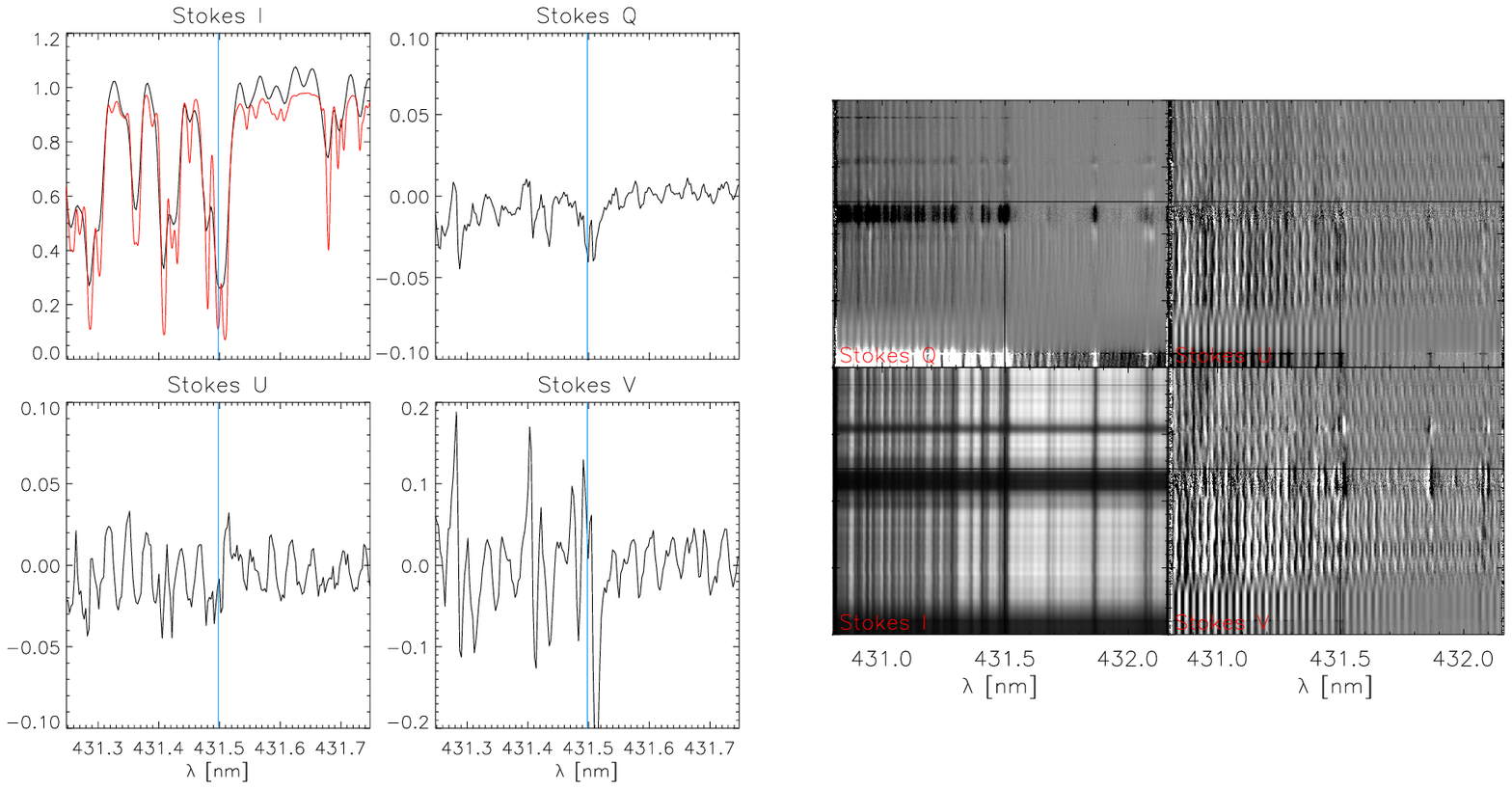}}}
  \caption{As in Fig.~\ref{specexam_426} but for \ion{Ti}{II} at 431.5\,nm.\label{specexam_431}}
\end{figure*}
\begin{figure*}
\centerline{\resizebox{17.6cm}{!}{\includegraphics{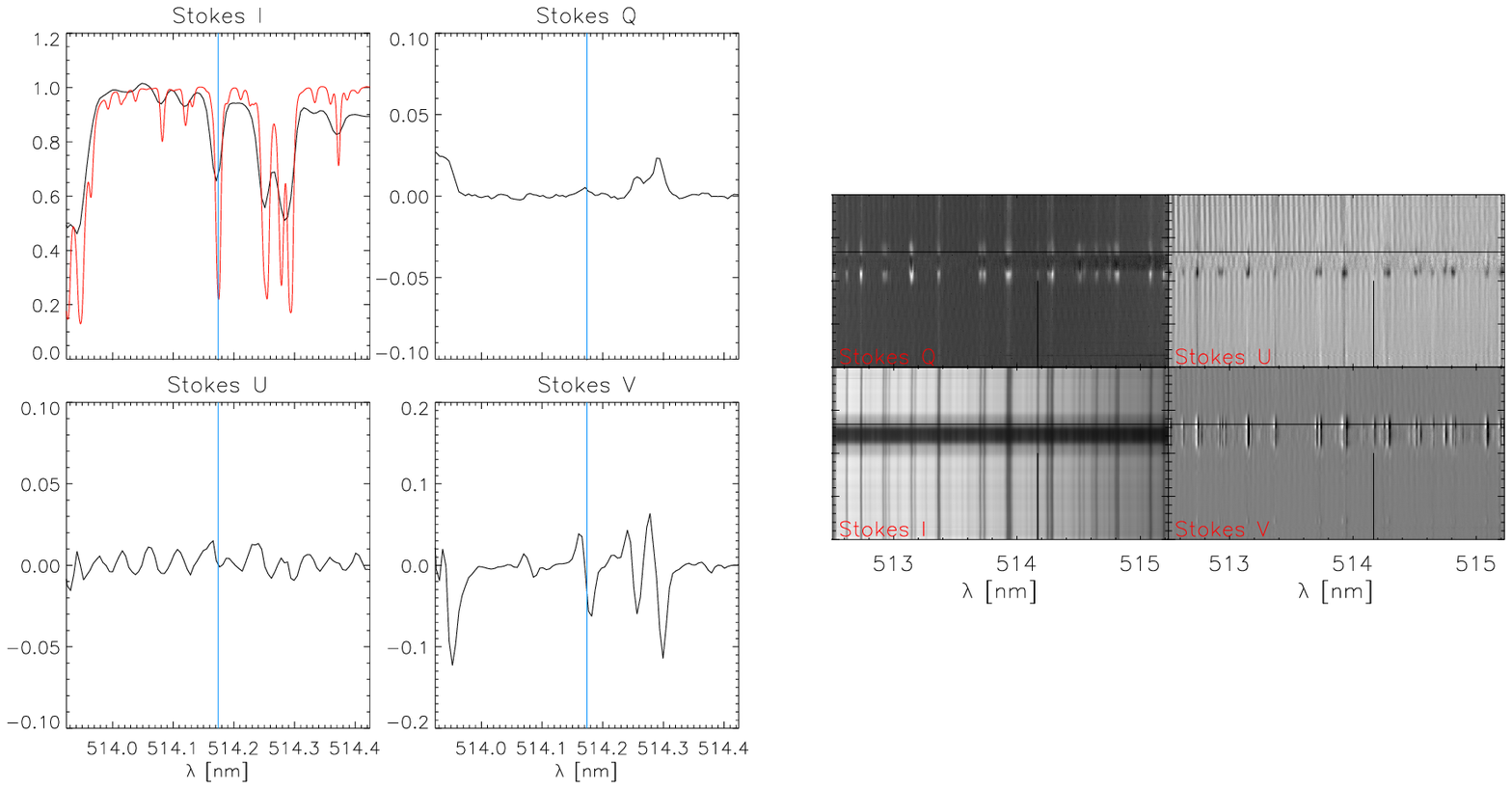}}}
\caption{Same as Fig.~\ref{specexam_426} for \ion{Fe}{I} at 514.2\,nm. \label{specexam_514}}
\end{figure*}

Corresponding example spectra from the observations are shown in Figs.~\ref{specexam_426} to \ref{specexam_514}. The \ion{Ti}{ii} line at 431.5\,nm is partly blended with other Zeeman-sensitive lines. The residual fringe amplitude of still several percent after removal of the strongest interference fringes makes any evaluation of the data at 425.8\,nm and 431.5\,nm unreliable. The observed peak Stokes $V$ polarization amplitudes of both lines of about 0.1 are sufficient for the inference of telescope properties as demonstrated with the \ion{Ca}{i} line at 526.2\,nm that has a maximal Stokes $V$ amplitude of 0.085. The \ion{Fe}{I} line at 514.2\,nm is well isolated in the spectrum (Figs.~\ref{fig_synth6} and \ref{specexam_514}) and does not suffer strongly from interference fringes in the observations, but its wavelength range was already covered by the \ion{Ca}{i} line at 526\,nm, so we did not evaluate these data.
\section{Transition parameters and polarization amplitudes \label{app_transi}}
Tables \ref{table_wolp} and \ref{table_wslp} contain the transition parameters and polarization amplitudes from the synthesis with 1500\,G for all lines without and with small intrinsic linear polarization, respectively. All spectral lines for which the maximal Stokes $V$ amplitude in the 1500\,G synthesis is about 8\,\% or more are suitable for the inference of telescope properties as demonstrated with the \ion{Ca}{i} line at 526.2\,nm. For lines with lower polarization amplitudes, the problem reduces to the question whether the crosstalk is large enough for the spurious signals to still be measured in the observations. For the case of the DST, the $V\rightarrow QU$ crosstalk can be huge with values of $T_{i4}/T_{44}$ above 1, but for other telescopes that number can be as small as a few percent. This implies that polarization signals at the $10^{-4} - 10^{-3}$ level have to be measured.

\begin{table*}
\caption{Spectral lines without linear polarization. Transition parameters, minimal line-core intensity, and maximal Stokes $V$ polarization amplitudes at 1500\,G. \label{table_wolp}}
\begin{tabular}{ccccccc|cc}
\hline\hline
Element & Ion. & $\lambda_0$  & Excitation & log (gf) & Transition & g$_{\rm eff}$  & min. $I$ & max. $|V/I|$ \\ 
        & state      & [\AA] & potential (eV)  &          &            &           \\ \hline
Cr &II& 4087.59 & 3.103 & -3.22 & 4D 0.5- 6D 0.5 & 1.667 & 0.77028 &  0.0090 \\ 
V & I & 4093.5 & 1.183 & -1.02 & 4P 0.5- 4D 0.5 & 1.333 & 0.92144  & 0.0028  \\ 
V & I & 4153.32 & 0.262 & -2.55 & 6D 0.5-4D 0.5 & 1.667  & 0.98063  & 0.0009  \\ 
Cu & I & 4248.95 & 5.076 & -0.976 & 4P 0.5- 4D 0.5 & 1.333 & 0.98631  & 0.0005\\ 
Mn & I & 4257.66 & 2.953 & -0.70 & 4D 0.5- 4P 0.5 & 1.333 & 0.18989  & 0.0277 \\ 
S &II& 4278.5 & 16.092 & -0.11 & 4P 0.5- 4D 0.5 & 1.333 & 1.00043  & 2.1e-8 \\ 
Ti &II& 4314.9 & 1.1609 & -1.104 & 4P 0.5- 4D 0.5 & 1.333 & 0.07868  & 0.0227\\ 
Fe &II& 4385.38 & 2.778 & -2.6 & 4P 0.5- 4D 0.5 & 1.333 & 0.12297 & 0.0216\\ 
Ti &II& 4407.67 & 1.221 & -2.617 & 2P 0.5- 4D 0.5 & 0.333 & 0.37181  & 0.0050\\ 
Ti &II& 4411.93 & 1.224 & -2.524 & 4P 0.5- 4D 0.5 & 1.333 & 0.32543  & 0.0215\\
S &II& 4456.38 & 15.848 & -0.55 & 4D 0.5- 4P 0.5 & 1.333& 1.00015  & 9.5e-09 \\ 
Cr &II& 4588.20 & 4.071 & -0.64 & 4D 0.5- 6F 0.5 & 0.333  & 0.18781  & 0.0054\\ 
V & I & 4626.48 & 1.043 & -1.24 & 4D 0.5- 4P 0.5&1.333 & 0.93216  & 0.0026 \\ 
V & I & 4757.48 & 2.029 & -0.29 & 6G 1.5- 6F 0.5 & 0.167 & 0.93489  & 0.0003\\ 
Co & I & 5247.93 & 1.785 & -2.08 & 4P 0.5- 4D 0.5 & 1.333 & 0.75913  & 0.0112\\
Mn & I & 5292.87 & 3.383 & -2.90 & 4P 0.5- 4D 0.5 & 1.333  & 0.99262  & 0.0003\\
S &II& 5473.62 & 13.584 & -0.226 & 4P 0.5- 4D 0.5 & 1.333 & 1.00030  & 3.3e-07\\ 
C &II& 5818.31 & 22.528 & -1.464 & 4D 0.5- 4P 0.5 & 1.333  & 0.99982  & 7.6e-17 \\ 
V & I & 6008.67 & 1.183 & -2.34 & 4P 0.5- 4D 0.5 & 1.333  & 0.99551  & 0.0002\\ 
V & I & 6111.67 & 1.043 & -0.715 & 4D 0.5- 4P 0.5 & 1.333  & 0.80069  & 0.0101\\
Co & I & 6117.00 & 1.785 & -2.49 & 4P 0.5- 4D 0.5 & 1.333 & 0.89543  & 0.0058\\
Fe &II& 6149.23 & 3.889 & -2.8 & 4D 0.5- 4P 0.5 & 1.333 & 0.54384  & 0.0186 \\ 
S &II& 6397.99 & 14.154 & -0.791 & 4D 0.5- 4P 0.5 & 1.333 & 0.99974  & 2.3e-08\\ 
N & I & 6646.50 & 11.750 & -1.539 & 4D 0.5- 4P 0.5 & 1.333  & 1.00055  & 4.2e-06 \\ 
C &II& 6787.21 & 20.701 & -0.377 & 4P 0.5- 4D 0.5 & 1.333& 1.00038  & 7.7e-12 \\ 
F & I & 6870.22 & 12.751 & -0.27 & 4P 0.5- 4D 0.5 & 1.333  & 1.00011  & 5.3e-09 \\ 
Co & I & 6872.40 & 2.008 & -1.85 & 4P 0.5- 4D 0.5 & 1.333 & 0.76954  & 0.0141\\ 
Cl & I & 8428.25 & 9.029 & -0.29 & 4P 0.5- 4D 0.5 & 1.333 & 0.99912  & 6.9e-05\\ 
N & I & 8703.25 & 10.326 & -0.310 & 4P 0.5- 4D 0.5 & 1.333  & 0.97682  & 0.0009\\
P & I & 10596.90 & 6.935 & -0.24 & 4P 0.5- 4D 0.5 & 1.333   & 0.94471  & 0.0033 \\ 
N & I & 11294.26 & 11.750 & -0.531 & 4D 0.5- 4P 0.5 & 1.333 & 0.99809 & 7.5e-05 \\
\end{tabular}
\end{table*} 

\begin{table*}
\caption{Spectral lines with small linear polarization. Transition parameters and maximal Stokes $QUV/I$ polarization amplitudes at 1500\,G.\label{table_wslp}}
\begin{tabular}{ccccccc|ccc}
\hline\hline
Elem. & Ion. & $\lambda_0$             & Excit. & log (gf) & Transition & g$_{\rm eff}$ & max. $|Q/I|$ & max. $|U/I|$ & max. $|V/I|$\\ 
        & state      & [\AA] & pot. (eV)  &          &            &                       \\ \hline
Ti & I & 4005.95 & 2.103 & -0.53 & 5F 3.0- 5H 4.0 & 0.375  & 0.0002 & 2.7e-06 & 0.0085\\ 
Fe & I & 4007.27 & 2.759 & -1.276 & 3G 3.0- 3F 2.0 & 0.500 & 0.0249 & 0.0147 & 0.0874\\
Fe& I & 4017.08 & 2.759 & -1.992 & 3G 3.0- 3D 2.0 & 0.333  & 0.0026 & 0.0003 & 0.0398\\
He & I & 4026.19 & 20.96 & -1.453 & 3P 1.0- 3D 1.0 & 1.000  & 2.3e-09 & 8.1e-15 & 8.0e-08\\
Fe & I & 4076.22 & 3.071 & -1.99 & 3P 1.0- 3D 1.0 & 1.000 & 0.0084 & 0.0020 & 0.0896\\ 
Fe & I & 4109.80 & 2.845 & -0.940 & 3P 1.0- 3D 1.0 & 1.000 & 0.0056 & 0.0064 & 0.0738 \\
Co & I & 4132.14 & 1.049 & -2.82 & 2F 2.5- 4D 2.5 & 1.114  & 0.0012 & 3.9e-05 & 0.0217\\ 
Al &II& 4227.95 & 15.062 & -1.709 & 3D 2.0- 3F 2.0 & 0.917  & 1.5e-09 & 7.8e-16 & 2.4e-08\\
Fe & I & 4229.51 & 3.274 & -1.628 & 3D 1.0- 3P 1.0 & 1.000  & 0.0106 & 0.0029 & 0.0906\\
Cr & I & 4232.23 & 4.207 & -0.50 & 3D 2.0- 3G 3.0 & 0.333 & 9.8e-05 & 7.5e-07 & 0.0036\\ 
Fe &II& 4296.57 & 2.704 & -2.9 & 4P 1.5- 4F 2.5 & 0.500  & 0.0071 & 0.0016 & 0.0519\\ 
Cl &II& 4304.04 & 15.712 & -0.68 & 3D 1.0- 3P 1.0 & 1.000  & 2.5e-12 & 1.1e-20 & 4.5e-11 \\ 
Sc &II& 4305.71 & 0.595 & -1.30 & 3F 2.0- 3D 2.0 & 0.917 & 0.0066 & 0.0013 & 0.0806 \\ 
Cr & I & 4312.47 & 3.113 & -1.37 & 3F 3.0- 3H 4.0 & 0.375& 0.0003  & 4.5e-06 & 0.0055  \\
Fe & I & 4367.90 & 1.608 & -2.886 & 3F 2.0- 5G 2.0 & 0.500 & 0.0004 & 0.0002 & 0.0554 \\
Cr & I & 4410.96 & 2.983 & -1.22 & 3H 5.0- 5F 4.0 & 0.400 & 0.0001 & 2.9e-06 & 0.0097\\ 
Fe & I & 4422.57 & 2.845 & -1.115 & 3P 1.0- 3D 1.0 & 1.000& 0.0085 & 0.0082 & 0.0784 \\
Cr & I & 4429.92 & 3.556 & -0.67 & 3D 1.0- 3P 1.0 & 1.000 & 0.0013 & 6.8e-05 & 0.0245 \\ 
Ca & I & 4435.69 & 1.886 & -0.519 & 3P 1.0- 3D 1.0 & 1.000 & 0.0078 & 0.0064 & 0.0735 \\
He & I & 4471.48 & 20.964 & -1.036 & 3P 1.0- 3D 2.0 & 1.000 & 1.9e-08 & 1.4e-13 & 1.4e-07 \\
Ni & I & 4513.0 & 3.706 & -1.47 & 3D 2.0- 3F 2.0 & 0.917 & 0.0021 & 0.0002 & 0.0337\\ 
Cr &II& 4539.59 & 4.0423 & -2.53 & 2F 2.5- 4D 2.5 & 1.114  & 0.0009 & 3.2e-05 & 0.0168\\
Al &II& 4589.67 & 15.062 & -1.608 & 3D 2.0- 3F 2.0 & 0.917  & 1.5e-09 & 9.0e-16 & 2.3e-08\\
P &II& 4626.70 & 12.812 & -0.32 & 3D 2.0- 3F 2.0 & 0.917  & 6.5e-09 & 1.5e-14 & 9.6e-08\\ 
Cr & I & 4698.94 & 3.079 & -1.44 & 3G 3.0- 3D 2.0 & 0.333  & 0.0001 & 1.1e-06 & 0.0047 \\
Ti & I & 4722.61 & 1.053 & -1.33 & 3P 1.0- 3D 1.0 & 1.000 & 0.0020 & 0.0001 & 0.0380\\
C & I & 4738.21 & 7.946 & -3.115 & 3D 1.0- 3P 1.0 & 1.000 & 7.0e-05 & 3.4e-07 & 0.0014\\
Cr & I & 4767.27 & 3.556 & -1.02 & 3D 2.0- 3F 2.0 & 0.917  & 0.0008 & 2.0e-05 & 0.0115 \\
Cl &II& 4778.91 & 17.086 & -0.35 & 3P 1.0- 3D 1.0 & 1.000 & 4.5e-13 & 5.6e-22 & 7.7e-12\\
Ni & I & 4808.87 & 3.706 & -1.41 & 3D 2.0- 3G 3.0 & 0.333 & 0.0003 & 1.2e-05 & 0.0151\\
Fe & I & 4813.11 & 3.274 & -2.84 & 3D 1.0- 5D 1.0 & 1.000 & 0.0020 & 0.0002 & 0.0399\\
Si & I & 4823.32 & 4.930 & -2.33 & 3P 1.0- 3D 1.0 & 1.000 & 0.0016 & 0.0001 & 0.0315 \\
Fe &II& 4833.19 & 2.657 & -4.8 & 4H 5.5- 6F 4.5 & 0.455 & 0.0002 & 2.7e-06 & 0.0079\\ 
Sr & I & 4876.08 & 1.798 & -0.551 & 3P 1.0- 3D 1.0 & 1.000 & 7.8e-05 & 1.8e-07 & 0.0014 \\ 
Cl &II& 4922.15 & 15.714 & -0.59 & 3D 2.0- 3F 2.0 & 0.917& 1.9e-12 & 7.3e-21 & 2.7e-11 \\ 
Cl &II& 4924.25 & 15.626 & -1.54 & 3F 2.0- 3D 2.0 & 0.917  & 2.5e-13 & 1.2e-22 & 3.4e-12\\
P &II& 4927.20 & 12.791 & -0.68 & 3D 1.0- 3P 1.0 & 1.000 & 2.3e-09 & 2.4e-15 & 3.8e-08  \\ 
Fe & I & 4930.32 & 3.960 & -1.201 & 3D 1.0- 5D 1.0 & 1.000& 0.0097 & 0.0025 & 0.0875 \\
Cr & I & 4936.34 & 3.113 & -0.34 & 3F 3.0- 3H 4.0 & 0.375 & 0.0018 & 0.0002 & 0.0317 \\ 
Ti & I & 4941.57 & 2.160 & -1.01 & 3D 2.0- 3F 2.0 & 0.917 & 0.0005 & 7.5e-06 & 0.0075\\ 
Ti & I & 4997.09 & 0.000 & -2.056 & 3F 2.0- 3D 2.0 & 0.917 & 0.0035 & 0.0005 & 0.0620\\ 
Fe & I & 5021.59 & 4.256 & -0.677 & 5F 3.0- 5H 4.0 & 0.375  & 0.0036 & 0.0005 & 0.0441 \\ 
Fe & I & 5099.08 & 3.984 & -1.265 & 3F 2.0- 3D 2.0 & 0.917  & 0.0092 & 0.0021 & 0.0830\\ 
Fe & I & 5141.74 & 2.424 & -2.238 & 3P 1.0- 3D 1.0 & 1.000  & 0.0162 & 0.0069 & 0.0916\\ 
Fe & I & 5207.94 & 3.635 & -2.40 & 3D 1.0- 3P 1.0 & 1.000  & 0.0022 & 0.0003 & 0.0465\\ 
Ca & I & 5261.71 & 2.521 & -0.73 & 3D 1.0- 3P 1.0 & 1.000  & 0.0120 & 0.0039 & 0.0845\\ 
Al &II& 5280.27 & 15.586 & -1.981 & 3P 1.0- 3D 1.0 & 0.917 & 1.8e-10 & 1.9e-17 & 3.0e-09 \\
Ti & I & 5300.01 & 1.053 & -1.47 & 3P 1.0- 3D 1.0 & 1.000 & 0.0016 & 8.7e-05 & 0.0307\\ 
C & I & 5300.87 & 8.640 & -2.622 & 3D 1.0- 3P 1.0 & 1.000 & 5.9e-05 & 3.0e-07 & 0.0011\\ 
Fe & I & 5341.02 & 1.608 & -1.953 & 3F 2.0- 3D 2.0 & 0.917 & 0.0072 & 0.0056 & 0.0778 \\
Ni & I & 5353.39 & 1.951 & -2.81 & 3P 1.0- 3D 1.0 & 1.000 & 0.0039 & 0.0006 & 0.0670\\ 
Ni & I & 5514.79 & 3.847 & -1.99 & 1F 3.0- 3P 2.0 & 0.500  & 0.0004 & 5.2e-06 & 0.0055 \\ 
Fe & I & 5667.66 & 2.609 & -2.94 & 3F 2.0- 3D 2.0 & 0.917  & 0.0064 & 0.0013 & 0.0790 \\ 
Fe & I & 5747.95 & 4.608 & -1.41 & 3F 3.0- 3H 4.0 & 0.375  & 0.0011 & 8.4e-05 & 0.0212\\ 
Fe & I & 6085.26 & 2.759 & -2.712 & 3G 3.0- 3D 2.0 & 0.333  & 0.0017 & 0.0002 & 0.0358\\
Fe & I & 6127.91 & 4.413 & -1.399 & 3F 3.0- 3H 4.0 & 0.375 & 0.0017 & 0.0002 & 0.0285 \\
\end{tabular}
\end{table*}
\begin{figure*}
\begin{center}
\resizebox{17.6cm}{!}{\includegraphics{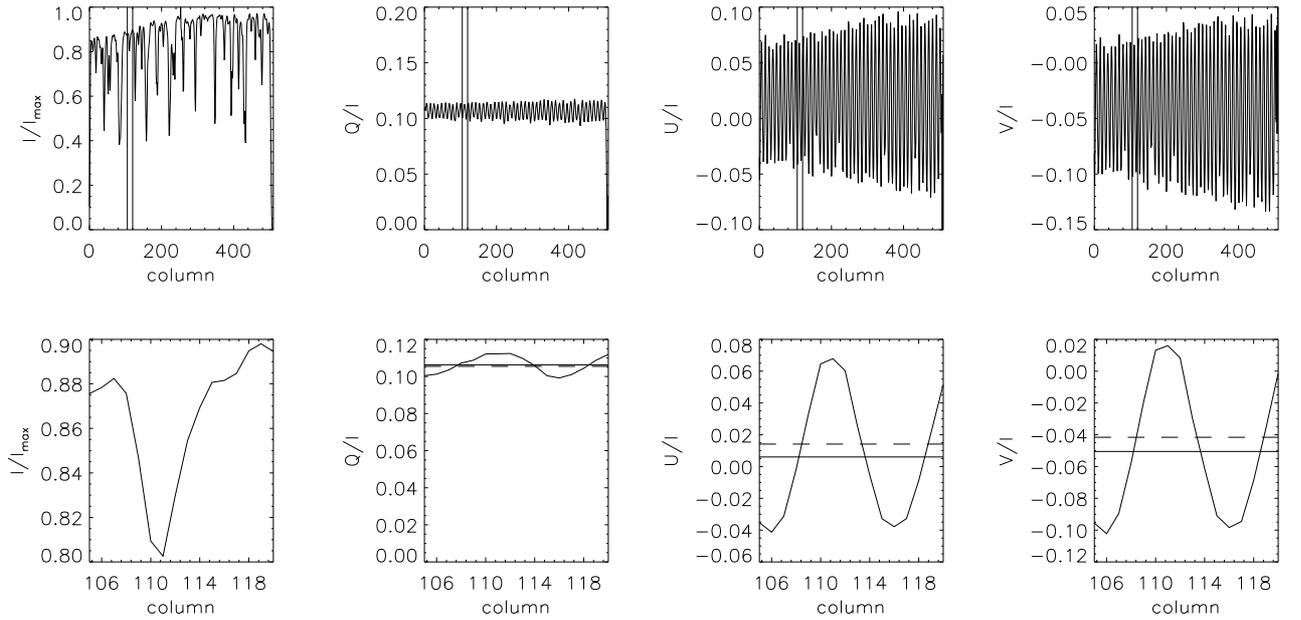}}\\
\caption{Determination of the $I\rightarrow QUV$ crosstalk at 459\,nm. Left to right: Stokes $I$, $Q/I$, $U/I$, and $V/I$ for a single, randomly picked profile. Top row: full spectrum. The two vertical lines indicate the location of the continuum wavelength window used in the determination of $I\rightarrow QUV$. Bottom row: spectrum inside the continuum wavelength window. The horizontal solid and dashed lines indicate the average value in the continuum window and the full spectrum, respectively.\label{figi2quv1}}
\end{center}
\end{figure*}
\begin{figure*}
\begin{center}
\resizebox{17.6cm}{!}{\includegraphics{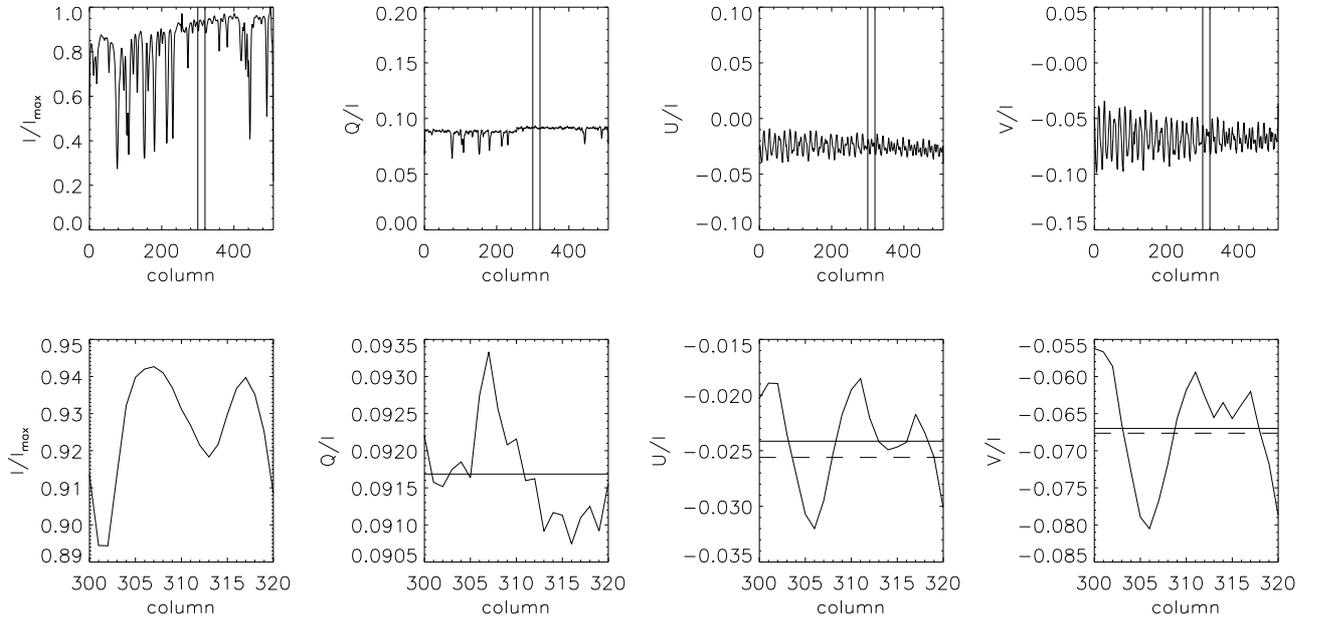}}\\
\caption{As in Figure \ref{figi2quv1} but at 526\,nm. \label{figi2quv2}}
\end{center}
\end{figure*}
\begin{figure*}
\begin{center}
\resizebox{17.6cm}{!}{\includegraphics{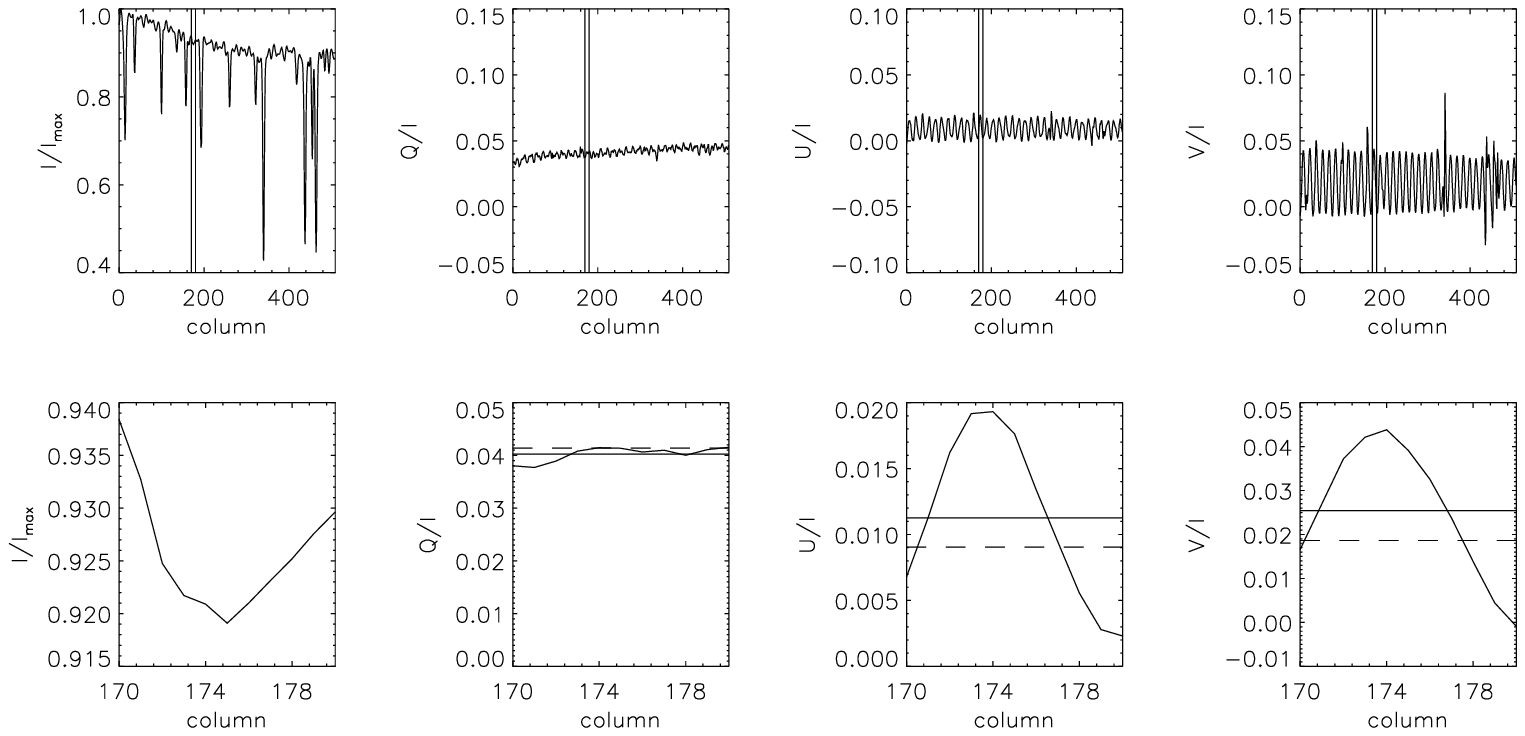}}\\
\caption{As in Figure \ref{figi2quv1} but at 615\,nm. \label{figi2quv3}}
\end{center}
\end{figure*}
\section{Error estimates for measurements \label{errestim}}
\subsection{Crosstalk from intensity to polarization $I\rightarrow QUV$\label{errori2quv}}
The crosstalk from intensity to polarization $I\rightarrow QUV$ was determined as part of the 2013 data reduction pipeline prior to fringe removal. A window of continuum wavelengths close to the spectral line of interest was specified (see Figs.~\ref{figi2quv1} to \ref{figi2quv3}). The extent of the window was chosen to be as large as possible without including solar spectral lines with genuine polarization signal. The $I\rightarrow QUV$ crosstalk was then determined as the average value of $QUV/I$ inside the continuum window. 

The fringe amplitude in the spectra can reach up to several percent but generally decreases with increasing wavelength (Figs.~\ref{figi2quv1} to \ref{figi2quv3}). Each time a complete fringe period is covered in the continuum wavelength window, its contribution to the average value of $QUV/I$ is zero. The maximum effect that fringes can cause happens when half a period of the fringe pattern is not compensated by the corresponding half with opposite sign. This residual contribution scales with the fringe period relative to the total extent of the continuum window. 

To estimate the possible error in the determination of the $I\rightarrow QUV$ crosstalk introduced by the fringe pattern, we randomly picked one individual profile from one map at 459, 526, and 615\,nm. We determined the value of $I\rightarrow QUV$ in the same continuum window as used in the data reduction and averaged it over the full wavelength range of the spectra for comparison (Table \ref{tabi2quv}). The difference between the two values is between 0.001 and 0.01 while Figs.~\ref{figi2quv1} to \ref{figi2quv3} show that the average value of $QUV/I$ is well recovered even in the presence of fringes in individual profiles. The fringe pattern showed a variation in time, space, and wavelength, whereas the $I\rightarrow QUV$ crosstalk values used in the fit of telescope parameters were an average of more than 100 profiles for each scan step. The error for a single value of $I\rightarrow QUV$ in, e.g., Fig. \ref{sign_flip} should thus be less than 1\,\%.

\begin{table*}
\begin{minipage}{8cm}
\caption{$I\rightarrow QUV$ for individual profiles.\label{tabi2quv}}
\centering
\begin{tabular}{c|cc}
\hline\hline
 & full  & continuum \cr
 & profile & window \cr\hline
$\lambda$&\multicolumn{2}{c}{458.8\,nm}\cr\hline
$Q/I$ & 0.105  &  0.106  \cr
$U/I$ & 0.014  & 0.006 \cr
$V/I$ & -0.042 &  -0.051 \cr\hline
$\lambda$&\multicolumn{2}{c}{526.2\,nm}\cr\hline
$Q/I$ &  0.088 &     0.092 \cr
$U/I$ &   -0.026&   -0.024\cr   
$V/I$ &   -0.068 &  -0.067\cr\hline   
$\lambda$&\multicolumn{2}{c}{614.9\,nm}\cr\hline
$Q/I$ & 0.041    & 0.041    \cr   
$U/I$ &   0.009   & 0.011     \cr
$V/I$ &     0.019 &     0.020 \cr\hline
\end{tabular}
\end{minipage}
\begin{minipage}{8cm}
\caption{Rms values of the fit residuals at 614.9\,nm.\label{tab_fitresiduals}}
\centering
\begin{tabular}{ccccc}
\hline\hline
$V\rightarrow Q$ & $V\rightarrow U$ & $I\rightarrow Q$ & $I\rightarrow U$ & $I\rightarrow V$ \cr \hline
0.093 & 0.074 & 0.022 & 0.026 & 0.024 \cr\hline
\end{tabular}
\end{minipage}
\end{table*}

\begin{figure*}
\centerline{\resizebox{15.cm}{!}{\includegraphics{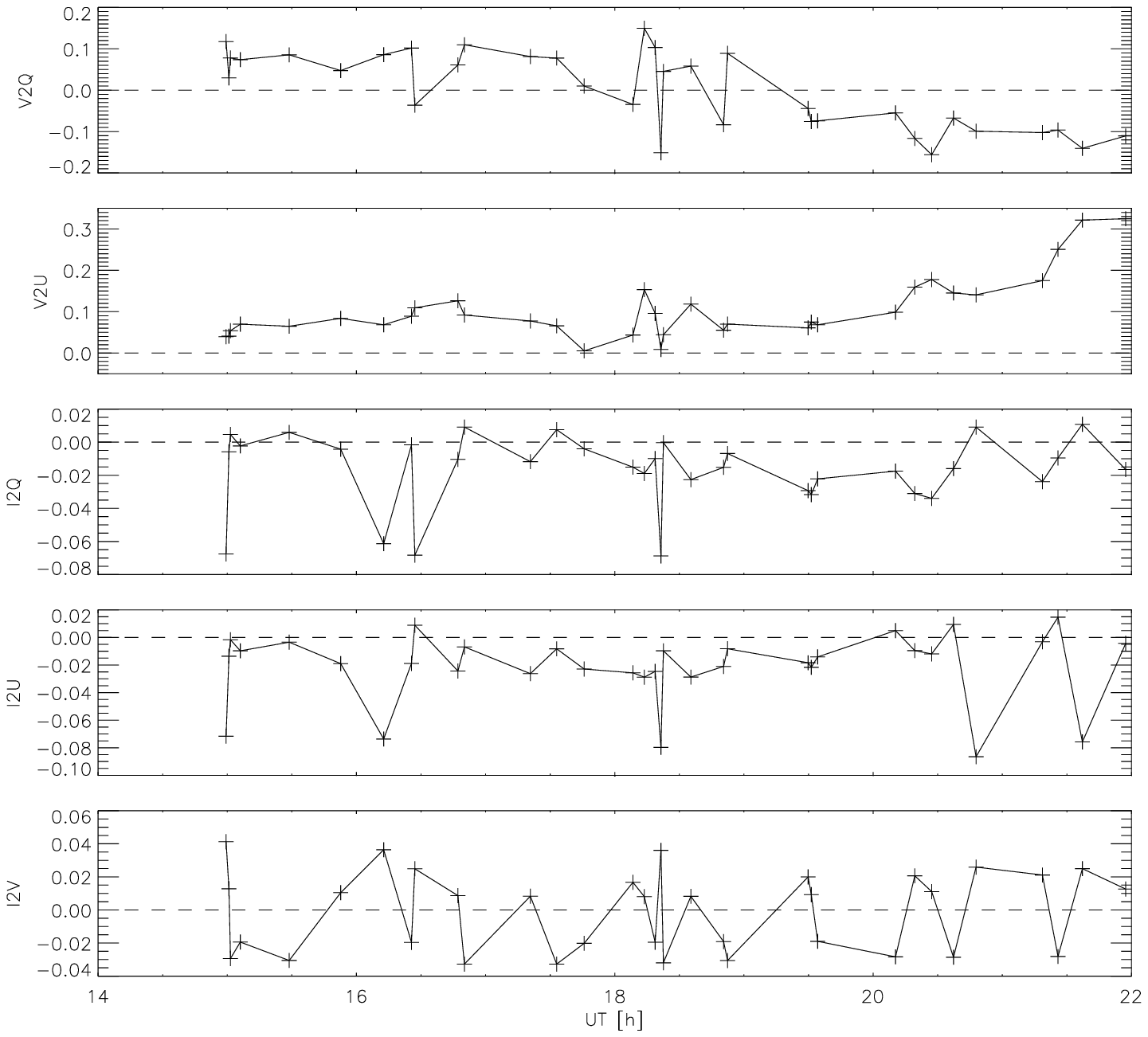}}}$ $\\$ $\\
\caption{Fit residuals at 614.9\,nm. Top to bottom: $V\rightarrow Q$, $V\rightarrow U$, $I\rightarrow QUV$. \label{fig_fitresiduals}}
\end{figure*}

\subsection{Crosstalk circular to linear polarization $V\rightarrow QU$\label{errorv2qu}}
It is somewhat more difficult to estimate the potential impact of the fringe pattern on the  $V\rightarrow QU$ crosstalk. The genuine polarization signal varies between individual profiles much more than the intensity crosstalk, in addition to the variable fringe pattern. Similar to the $I\rightarrow QUV$ crosstalk, the data points for $V\rightarrow QU$ were obtained by an average along the slit, but now over a variable number of pixels depending on the presence of significant polarization signal in the masks (see Fig.~\ref{masks}). 

As an estimate of the error in the $V\rightarrow QU$ crosstalk, we used the fit residuals at 614.9\,nm for all quantities used in the fit (see Fig.~\ref{fig_fitresiduals}), i.e., the difference between the observations and the final best-fit solution for the telescope model in $V\rightarrow QU$ and $I\rightarrow QUV$. The standard deviation of the residuals can be used as a first-order estimate of the random errors in the quantities although the residuals at this specific wavelength also show some more systematic variation, e.g., a sort of a linear trend in $V\rightarrow Q$ and an offset from zero in $V\rightarrow U$. Table \ref{tab_fitresiduals} lists the rms values of the fit residuals at 614.9\,nm. The rms of the $V\rightarrow QU$ residuals is about 8\,\%, while $I\rightarrow QUV$ fluctuates by about 2\,\%. The error of individual data points in $V\rightarrow QU$ used in the fit thus should be about 5\,\% when one considers the contribution of the systematic effects to the fluctuation around the mean value. 

As a last generic error estimate, we used the value of the $\chi^2$ at 614.9\,nm. If the reduced $\chi^2$ is assumed to follow a $\chi^2$ distribution, its value should be 
\begin{eqnarray}
\chi^2_{\rm reduced} = \frac{\chi^2}{N -f} \cdot \frac{1}{\sigma^2} \equiv 1 \,,
\end{eqnarray}
where $\chi^2 = \sum_i (OBS_i - FIT_i)^2$, $N$ is the number of data points, $f$ the number of degrees of freedom in the fit and $\sigma$ the error of one data point. 

With  $\chi^2 ({\rm 614.9\,nm}) = 4.808$, $N = 5 \cdot 34$ and $f = 5,$ one obtains $\sigma = 0.17$. This number is about twice as large as the previous estimates for $V\rightarrow QU$, which presumably indicates that the assumption of a well-behaved $\chi^2$ distribution is not fully valid.
\end{appendix}
\end{document}